\documentclass[aps,prd,showpacs,preprint,groupedaddress]{revtex4}

\usepackage[dvipdfmx]{graphicx}
\usepackage[FIGTOPCAP]{subfigure}

\usepackage{amsfonts}
\usepackage{amssymb}
\usepackage{amsmath}

\usepackage{ulem}
\usepackage{color}

\newcommand{\Slash}[1]{{\ooalign{\hfil/\hfil\crcr$#1$}}}

\begin{document}

\title{
  Partial restoration of chiral symmetry in the color flux tube
}

\author{Takumi~Iritani$^1$}
\email{iritani@yukawa.kyoto-u.ac.jp}
\author{Guido~Cossu$^2$} 
\author{Shoji~Hashimoto$^{2,3}$}
\affiliation{
  $^1$ Yukawa Institute for Theoretical Physics, \\
Kyoto University, Kitashirakawa-oiwake, Sakyo, Kyoto 606-8502, Japan}

\affiliation{
  $^2$ High Energy Accelerator Research Organization (KEK), \\
  Tsukuba, Ibaraki 305-0801, Japan}

\affiliation{
  $^3$ School of High Energy Accelerator Science, The Graduate University for Advanced Studies
(Sokendai), Tsukuba, Ibaraki 305-0801, Japan}

\date{\today}

\begin{abstract}
  Using the quark eigenmodes computed on the lattice with the
  overlap-Dirac operator,
  we investigate the spatial distribution of the chiral condensate
  around static color sources corresponding to quark-antiquark and
  three-quark systems.
  A flux structure of chromo fields appears in the presence of such color charges.
  The magnitude of the chiral condensate is reduced inside the color
  flux, which implies partial restoration of chiral symmetry inside hadrons.
  Taking a static baryon source in a periodic box as a toy model of nuclear matter, 
  we estimate the magnitude of the chiral symmetry restoration 
  as a function of baryon matter density.
\end{abstract}

\preprint{YITP-15-9, KEK-CP-319}

\maketitle

\section{Introduction}
The low-energy dynamics of Quantum Chromodynamics (QCD) is
characterized by two prominent properties, {\it i.e.}
chiral symmetry breaking and confinement.
In the QCD vacuum, chiral symmetry of quark fields is
spontaneously broken, as probed by an order parameter 
$\langle\bar{q}q\rangle$, the vacuum expectation value of the scalar 
density operator $\bar{q}q$.
Through the Banks-Casher relation \cite{Banks:1979yr}, this chiral
condensate $\Sigma=-\langle\bar{q}q\rangle$
can be related to the spectral density of the low-lying Dirac eigenvalues. 
In the presence of valence quarks, or color sources, some modification
of the {\it vacuum} is expected around them.
Strictly speaking, it is no longer the vacuum, {\it i.e.} the lowest
energy state of the system,
but we use this terminology having in mind an application to the study of finite
density QCD.
Although the static color sources as considered in this work, is only a
crude approximation to the finite density system, the study of the
vacuum modification may shed light on the states of finite
density QCD, which has been a subject of active research 
(see, for instance \cite{Fukushima:2010bq}).

The other interesting property of low-energy QCD is the confinement of
quarks, which is characterized by the linearly rising potential between
static color sources.
Putting a pair of static quark and anti-quark in the vacuum,
a color flux-tube emerges between them and leads to a linear
increase of the energy as a function of the separation.
This flux-tube structure has been observed in lattice QCD calculations 
by monitoring the action density or chromo-electric (or chromo-magnetic)
field \cite{Bali:1994de,Haymaker:1994fm,Cea:1995zt}.
We expect that such flux-tube structure is reflected in the low-lying
fermion eigenmodes, because the Dirac eigenmodes carry the information
of their background gauge field configuration.
Indeed, the QCD field-strength tensor can be reconstructed using the
fermion eigenmodes \cite{Gattringer:2002gn}.

In this paper we present a lattice study of the spatial distribution
of the chiral condensate in the presence of static color charges.
We consider quark-antiquark and three-quark systems represented by
Wilson loops to mimic the mesonic and baryonic states, respectively. 
We use the lattice data of the Dirac eigenmodes calculated on the
gauge configurations generated with 2+1-flavor dynamical overlap 
fermions \cite{Aoki:2012pma}.
With the overlap fermion formulation 
\cite{Neuberger:1997fp,Neuberger:1998wv},
chiral symmetry is exactly realized on the lattice,
which is important in the study of the low-lying Dirac eigenmodes, as
they are very sensitive to any small violations of chiral symmetry.
The lattice data used in this work have this nice property, and indeed
were successfully applied to the extraction of the chiral condensate
in the vacuum
\cite{Fukaya:2007fb,Fukaya:2007yv,Fukaya:2009fh,Fukaya:2010na}.

The organization of this paper is as follows.
In Section~\ref{sec:2}, we describe the method to construct 
the \textit{local chiral condensate} $\bar{q}q(x)$ by using the
overlap-Dirac eigenmodes, and show its distribution in the vacuum.
In Sections~\ref{sec:3} and \ref{sec:4}, we investigate the spatial
distribution of the local chiral condensate around the static color sources.
Section~\ref{sec:summary} is devoted to a summary.
Preliminary reports of this work are found in
\cite{Iritani:2013rla,Iritani:2014jqa,Iritani:2014fga}.

\section{Topological structure of the QCD vacuum}
\label{sec:2}

We investigate the topological structure of the non-perturbative QCD
vacuum in terms of the eigenmodes of the overlap-Dirac operator.
It preserves exact chiral symmetry, and the relation to the
topological charge of background gauge field configuration is
manifest, {\it i.e.} the index theorem, at least for smooth enough
backgrounds \cite{Niedermayer:1998bi}.

In this paper, we use the 2+1-flavor dynamical overlap-fermion
configurations generated by the JLQCD Collaboration
\cite{Aoki:2012pma}.
Their lattice volumes are $16^3\times 48$ and $24^3\times 48$
at a single inverse lattice spacing $a^{-1} = 1.759(10)$~GeV.
The dynamical quark masses are 
$m_{ud} = 0.015a^{-1}$ and $m_s = 0.080a^{-1}$. 
The global topological charge is fixed at $Q=0$ to avoid the problem
of divergent molecular dynamics force in the simulations
\cite{Fukaya:2006vs}.
It induces finite volume effects \cite{Aoki:2007ka},
which would not be significant for the relatively local observables 
considered in this study.
Most of the results are obtained on the larger lattice 
($24^3\times 48$) where the number of independent configurations is 50.

In the following we describe the profile of the low-lying eigenmodes
on these lattices.

\subsection{Local chiral condensate $\bar{q}q(x)$}

The massless overlap-Dirac operator is given by 
\cite{Neuberger:1997fp,Neuberger:1998wv}
\begin{equation}
  D_{\rm ov}(0) = m_0 \left[ 1 + \gamma_5 \mathrm{sgn} \ H_W(-m_0) \right],
  \label{eq:overlap-Dirac-operator}
\end{equation}
with the hermitian Wilson-Dirac operator $H_W(-m_0) = \gamma_5 D_W(-m_0)$.
Here, sgn denotes the matrix sign function.
Introducing the quark mass $m_q$,
the overlap-Dirac operator is modified as
\begin{equation}
  D_{\rm ov}(m_q) = \left( 1 - \frac{m_q}{2m_0} \right) D_\mathrm{ov}(0) + m_q.
\end{equation}
This form cancels $\mathcal{O}(a)$ discretization effects, together
with a proper rotation of the fermion fields in the observables.

We define the eigenfunction $\psi_\lambda(x)$ associated with an
eigenvalue $\lambda$ of the massless overlap-Dirac operator
\begin{equation}
  \label{eq:eigeneq}
  D_{\rm ov}(0) \psi_\lambda(x)=\lambda\psi_\lambda(x),
\end{equation}
where the eigenfunction $\psi_\lambda(x)$ is normalized as 
$\sum_x \psi_\lambda^\dagger(x) \psi_\lambda(x) = 1$.
Using them we may expand the ``local chiral condensate''
$\bar{q}q(x)$ in terms of the eigenmodes,  {\it i.e.}
\begin{equation}
  \bar{q}q(x) = - \sum_\lambda 
  \frac{
    \psi_\lambda^\dagger(x)\psi_\lambda(x)
  }{
    m_q + (1 - \frac{m_q}{2m_0})\lambda
  }
  \label{eq:localChiralCondensate}
\end{equation}
for a valence quark mass $m_q$.
This relation represents a self-contracting fermion loop contribution from and
to the scalar density operator.
If the measured observables do not include other light quark fields to be contracted,
then the substitution (\ref{eq:localChiralCondensate}) is justified.
The correlation functions of $\bar{q}q(x)$ with the Wilson-loop are in
this class of observables.

The chiral condensate $\langle \bar{q}q \rangle$ is given by 
an ensemble average of $\bar{q}q(x)$ without insertions of other
operators. 
By averaging over space-time, this quantity is simply written in terms
of only the eigenvalues because of the normalization condition for
$\psi_\lambda(x)$.
Thus the relation between the chiral condensate and the spectral
density $\rho(\lambda)$ of the Dirac eigenvalues is established. 
In the chiral limit it reads $\Sigma=\pi\rho(0)$, 
{\it i.e.} the Banks-Casher relation \cite{Banks:1979yr}.

\subsection{Action and topological charge densities in terms of the
  Dirac eigenmode}
Since the gauge field strength tensor $F_{\mu\nu}$ is defined through
the covariant derivative $D_\mu$ as $F_{\mu\nu}=[D_\mu,D_\nu]$, it can
also be related to the Dirac operator \cite{Gattringer:2002gn}.
Here we briefly reproduce the derivation.

The square of the Dirac operator $\Slash{D}\equiv\gamma_\mu D_\mu$
is decomposed as
\begin{equation}
  [\Slash{D}(x)]^2 = 
  \sum_\mu D_\mu^2(x) + 
  \sum_{\mu < \nu} \gamma_\mu \gamma_\nu F_{\mu\nu}(x).
  \label{eq:Dslash2}
\end{equation}
By multiplying $\gamma_\mu\gamma_\nu$ and taking a trace with respect
to the Dirac indices, the field strength tensor is expressed as
\begin{equation}
  F_{\mu\nu}(x) = - \frac{1}{4} \mathrm{tr}
  \left[ \gamma_\mu \gamma_\nu \Slash{D}^2(x) \right].
  \label{eq:Fmunu}
\end{equation}
Therefore, by expanding the Dirac operator in terms of its eigenvectors
$\psi_\lambda(x)$, an expansion of the field strength is obtained:
\begin{equation}
  F_{\mu\nu}(x) = \sum_\lambda \lambda^2 f_{\mu\nu}(x)_\lambda,
  \;\;
  f_{\mu\nu}(x)_\lambda \equiv 
  \frac{i}{2} \psi_\lambda^\dagger(x)\gamma_\mu\gamma_\nu\psi_\lambda(x).
  \label{eq:field_strength}
\end{equation}
Using this decomposition, the action and topological charge densities
are expressed as
\begin{eqnarray}
  \rho(x) = \mathrm{tr_c}[F_{\mu\nu} F_{\mu\nu}]
  &=&  \mathrm{tr_c} \sum_{\lambda,\lambda'} \lambda^2 \lambda'^2 
  f_{\mu\nu}(x)_\lambda
  f_{\mu\nu}(x)_{\lambda'},
  \label{eq:action_dens} \\
  q_{\mathrm{top}}(x) = \mathrm{tr_c}[F_{\mu\nu} \tilde{F}_{\mu\nu}]
  &=&  \mathrm{tr_c} \sum_{\lambda,\lambda'} \lambda^2 \lambda'^2 
  f_{\mu\nu}(x)_\lambda \tilde{f}_{\mu\nu}(x)_{\lambda'},
  \label{eq:topolo_dens}
\end{eqnarray}
respectively.
Here, $\mathrm{tr}_c$ denotes the trace with respect to the color
indices and 
$\tilde{f}_{\mu\nu}(x)_\lambda =
\frac{1}{2}\varepsilon_{\mu\nu\rho\sigma}f_{\rho\sigma}(x)_\lambda$.

\begin{figure}[tb]
  \centering
  \includegraphics[width=0.45\textwidth,clip]{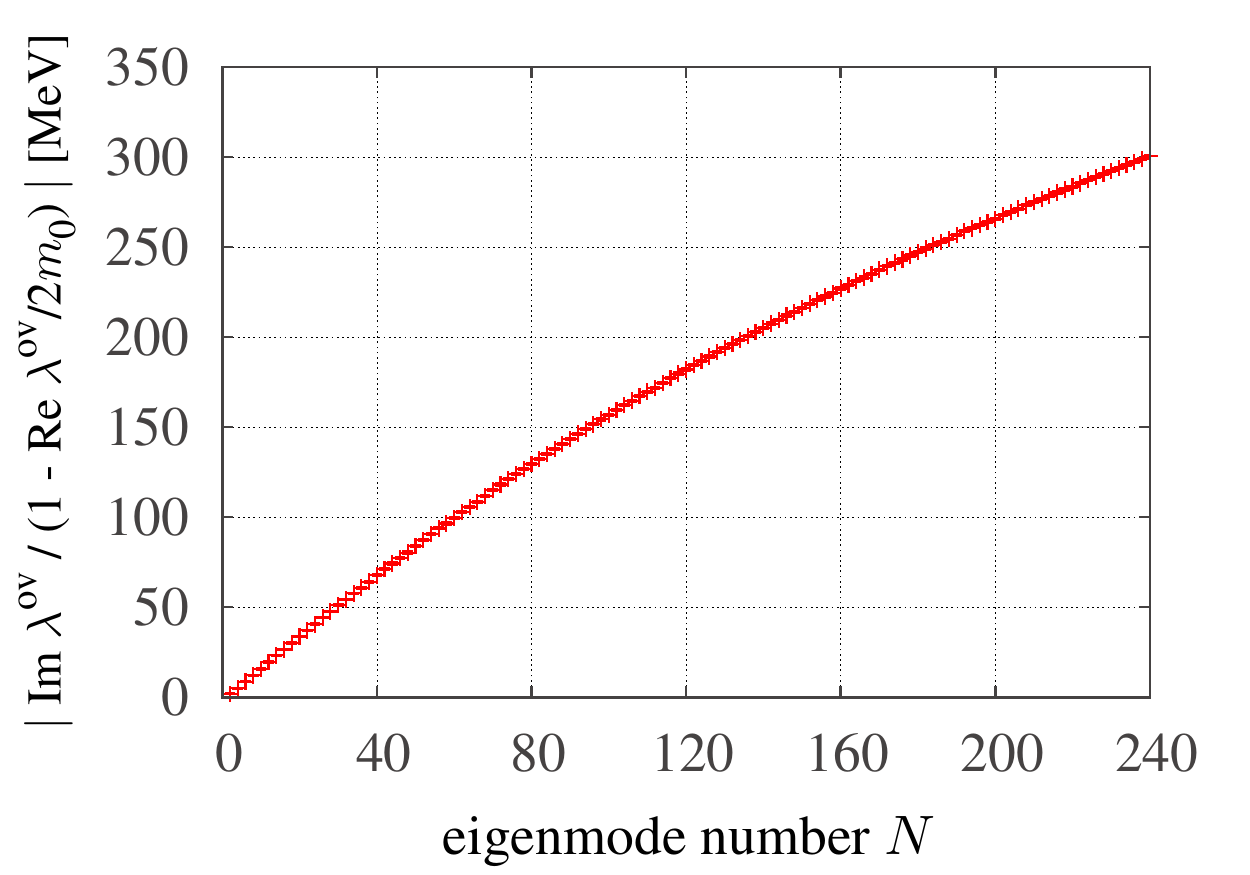}
  \caption{
    \label{fig:eigval}
    Eigenvalue (in MeV) of the low-lying eigenmodes on the
    $24^3\times 48$ lattice.
    By including 240 eigenmodes, we can cover the range of
    $\lambda\lesssim$ 300~MeV.
  }
\end{figure}

So far the expressions are exact, but in the numerical studies we introduce a truncation of the
summation over the eigenmodes. 
This truncation acts as a filter to cut UV fluctuations above $\lambda_\mathrm{max}$.
On the ensembles of $16^3\times 48$ and $24^3\times 48$ lattices, we
calculated 160 and 240 pairs of eigenvalues and eigenvectors of
$D_{\mathrm{ov}}$, respectively.
Then, the eigenvalues after correcting the $O(a)$ effect,
$\mathrm{Im}\ \lambda/(1-\mathrm{Re}\ \lambda/2m_0)$,
cover the region between $\pm 300$~MeV, as shown in
Figure~\ref{fig:eigval}.
In the measurements of the correlation between 
$\bar{q}q(x)$
and
the Wilson loops, we monitor the dependence on the number $N$ of the
eigenmodes included and confirm that the results saturate at least
above 200~MeV.
Some examples will be shown later.

\begin{figure*}[tb]
  \subfigure[\ local chiral condensate]{
    \includegraphics[width=0.45\textwidth,clip]{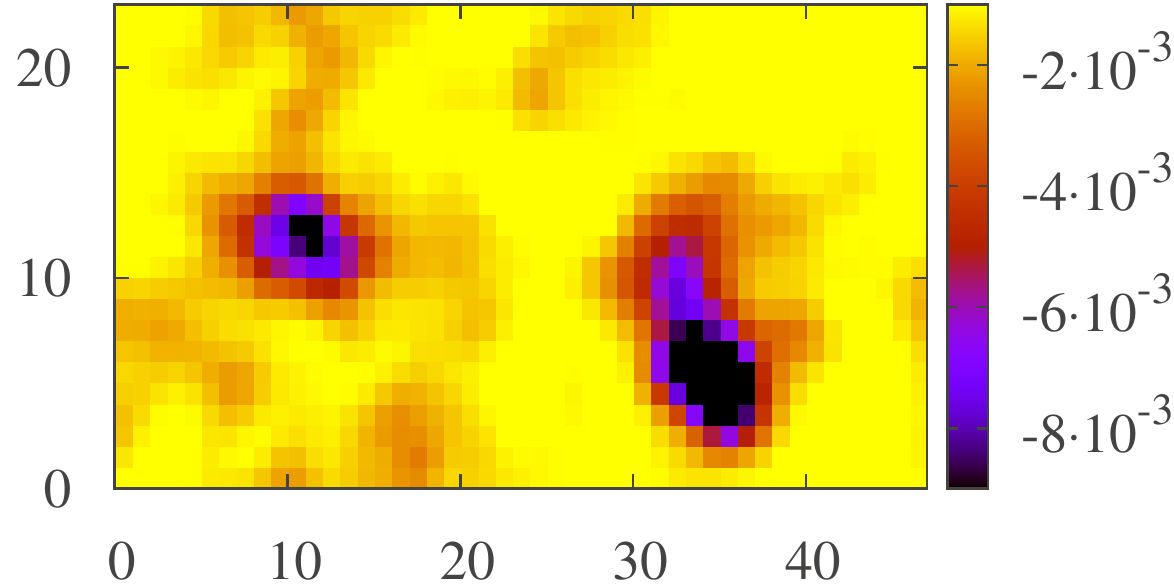}
  }
  \subfigure[\ action density]{
    \includegraphics[width=0.45\textwidth,clip]{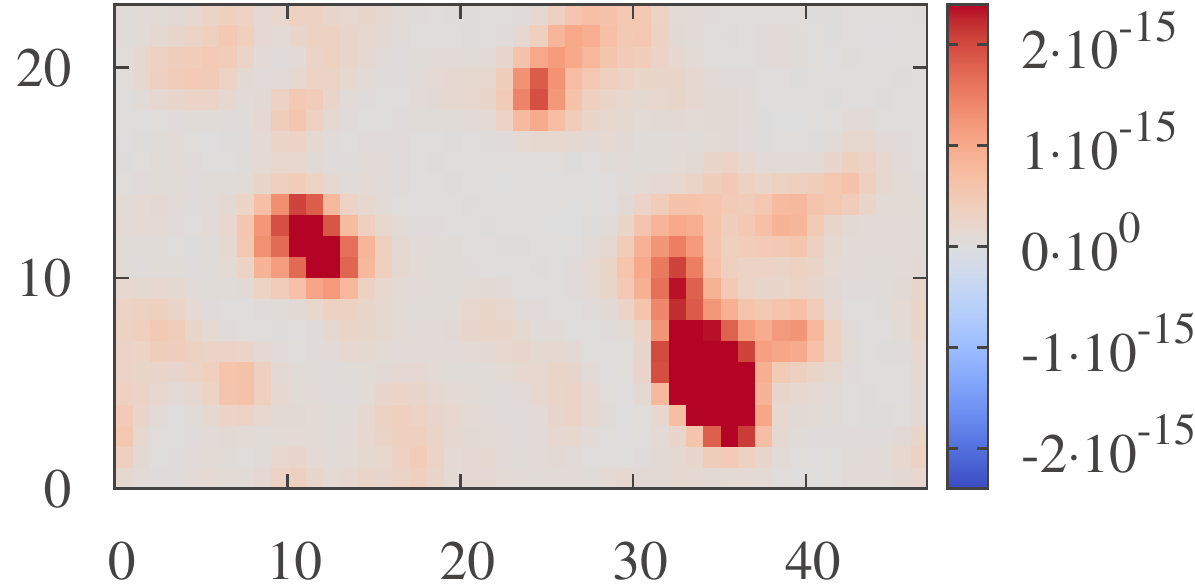}
  }
  \subfigure[\ topological charge density]{
    \includegraphics[width=0.45\textwidth,clip]{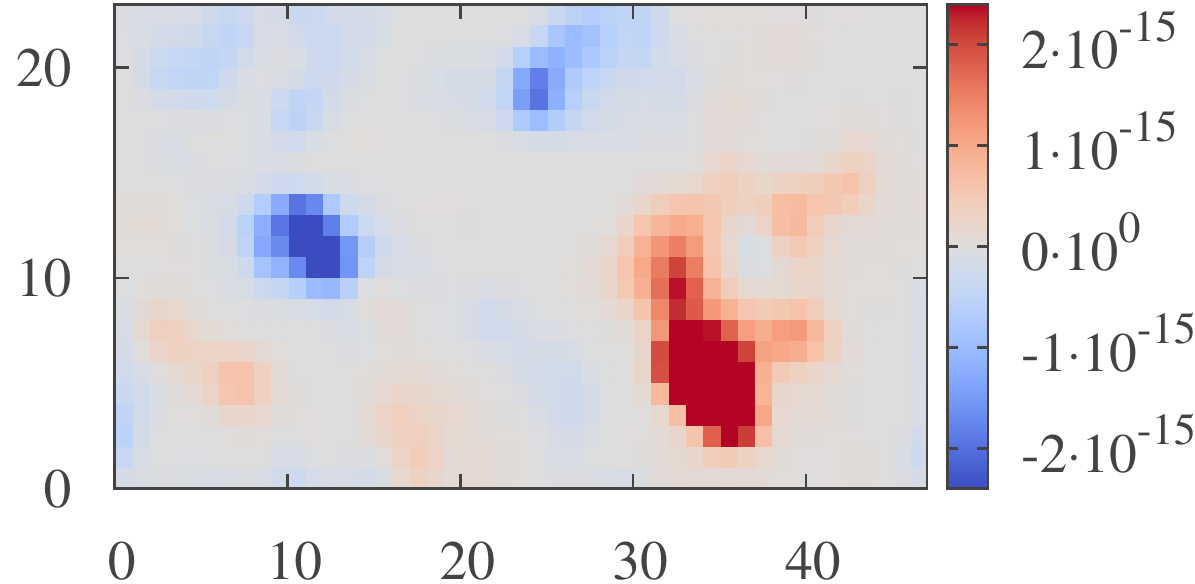}
  }
  \caption{
    \label{fig:snapshot_local_densities}
    Snapshots of (a) the local chiral condensate, 
    (b) the action density and (c) the topological charge 
    distributions observed with the sum of 20 lowest-lying eigenmodes.
    These pictures show the same $T$-$X$ slice of a $24^3 \times 48$
    lattice on a representative gauge configuration.
      The chiral condensate local fluctuations are correlated
      with the local topological charge measurements.
  }
\end{figure*}

Before presenting the results, we show some snapshots of the
eigenmodes. 
The index theorem dictates that exact zero-modes are associated with
topological excitations of the gauge field.
This suggests that the near-zero modes are superpositions of such
local topological objects.
Using a truncation at $N=20$, we visualize the low-mode contributions
to the local chiral condensate $\bar{q}q(x)$
(\ref{eq:localChiralCondensate}), 
the action density $\rho(x)$ (\ref{eq:action_dens}) and 
the topological charge density $q_{\mathrm{top}}(x)$
(\ref{eq:topolo_dens})
in the panels (a), (b) and (c) of
Figure~\ref{fig:snapshot_local_densities},
respectively.
They show tomographic images on a certain $T$-$X$ slice of the
four-dimensional lattice 
extracted from a given gauge configuration of size $24^3\times 48$.

As one can see in Fig.~\ref{fig:snapshot_local_densities} (a),
the local condensate $\bar{q}q(x)$ forms a cluster structure.
At the same space-time points of the cluster, the action density shows
peaks (the panel (b)).
More importantly, the topological charge density has positive and
negative islands stretching over several lattice spacings at the same
space-time points.
Such observation is not new; indeed, there are lattice studies using
the overlap-Dirac operator \cite{Ilgenfritz:2007xu,Ilgenfritz:2008ia}
showing the similar profile of the low-lying eigenmodes.

\section{Chiral condensate in Quark-Antiquark System}
\label{sec:3}

In the presence of color charges, there appears a flux-tube of
chromo-electric fields,
which has been observed on the lattice by measuring the spatial
distribution of the field strength tensor
\cite{Bali:1994de,Haymaker:1994fm,Cea:1995zt}.
In this section, we investigate the spatial distribution of the local
chiral condensate $\bar{q}q(x)$ (\ref{eq:localChiralCondensate})
around the static color sources.
Previously, a related analysis has been made, but on a single
color source, {\it i.e.} a Polyakov line
\cite{Feilmair:1988js,Sakuler:1992qx,Faber:1993sw},
or at finite temperature
where the flux-tube is expected to be suppressed
\cite{Chagdaa:2006zz}.

\subsection{Partial restoration of the chiral symmetry in the flux-tube}

\begin{figure}[tb]
  \includegraphics[width=0.45\textwidth,clip]{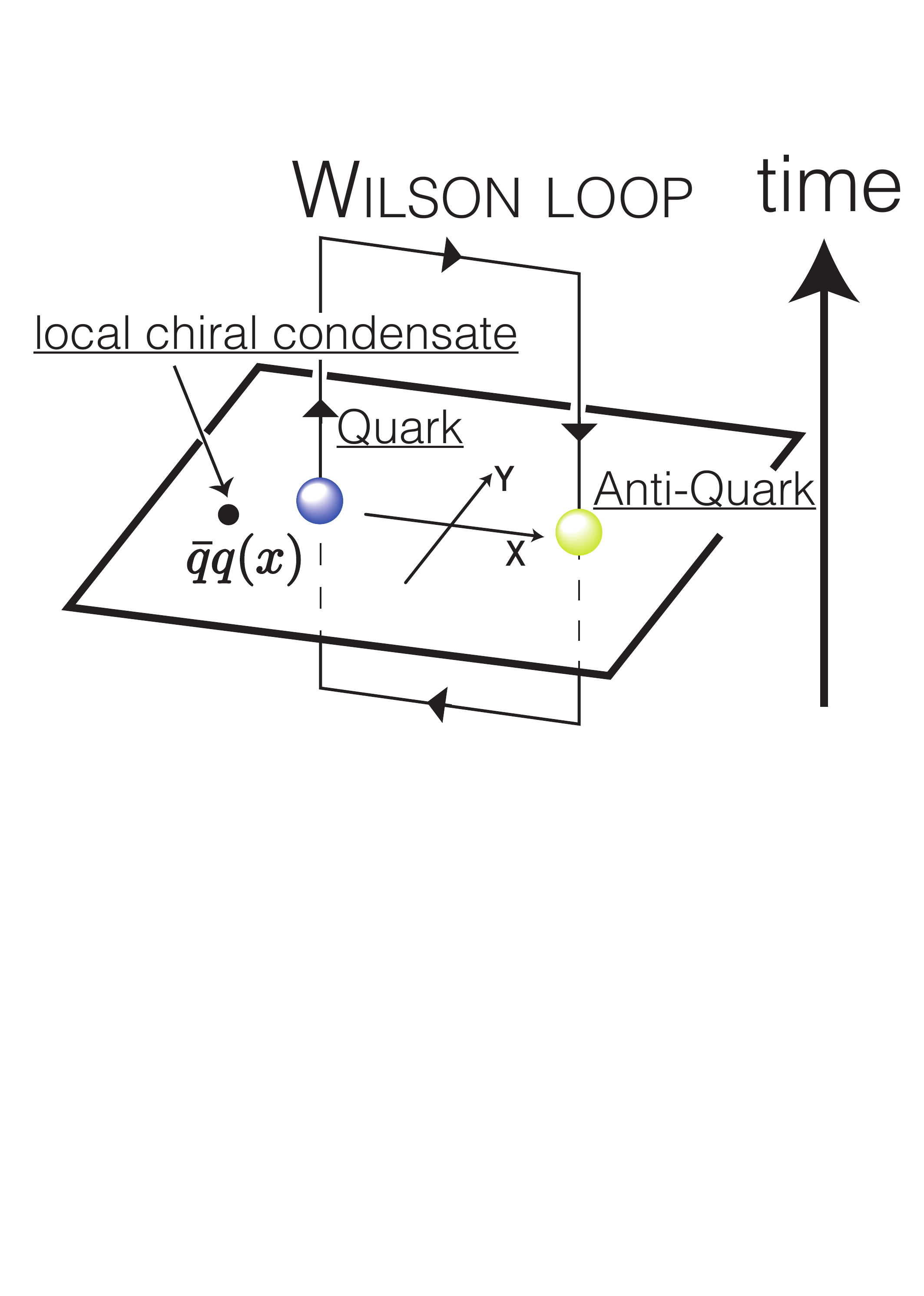}
  \caption{
    \label{fig:flux-tube-measurement}
    Schematic picture of the flux-tube measurement.
    Static quark and antiquark are located at $(R/2,0)$ and $(-R/2,0)$
    on the $XY$-plane.
  }
\end{figure}

We investigate the spatial distribution of the local chiral condensate
$\bar{q}q(x)$ around the static color sources by calculating a
correlation 
\begin{equation}
    \langle\bar{q}q(\vec{x})\rangle_W 
    \equiv \frac{\langle\bar{q}q(\vec{x}) W(R,T)\rangle}{\langle W(R,T)\rangle}
    - \langle \bar{q}q \rangle,
  \label{eq:diffLocalChiral}
\end{equation}
where $W(R,T)$ denotes a Wilson-loop of size $R \times T$.
It represents a pair of static quark and anti-quark separated by a
distance $R$. 
The origin of the coordinate is chosen at the center of the loop,
which stretches along the $X$-axis,
and the $Y$ and $Z$-axes correspond to the transverse directions.
Figure~\ref{fig:flux-tube-measurement} shows a schematic picture of the measurement.

As mentioned in the previous section, we truncate the sum over the
eigenmodes in (\ref{eq:localChiralCondensate}) at the $N$-th
eigenvalue and denote the corresponding local condensate as
$\bar{q}q^{(N)}(x)$.
In the final analysis, we chose $N=160$, after confirming that the
result is unchanged once a sufficient number of low-lying modes are included.

Reflecting the ultraviolet divergences of the scalar density operator,
the expectation value of $\bar{q}q^{(N)}(x)$ contains quadratic and
logarithmic divergences.
The strong quadratic divergence is associated with a mixing with
the identity operator and has the form $m_q/a^2$.
Because of the exact chiral symmetry of the overlap-Dirac operator,
the strongest divergence of $1/a^3$ is absent and the leading term is
of $1/a^2$ and proportional to $m_q$.
Since the truncation at a fixed mode number $N$ can be considered as
a certain regularization scheme, the regularized operator
$\bar{q}q^{(N)}(x)$ can be parametrized as
\begin{equation}
  \bar{q}q^{(N)} = 
  \bar{q}q^{\rm (subt)} + c_1^{(N)}m_q/a^2 + c_2^{(N)}m_q^3
  \label{eq:subtractChiral}
\end{equation}
with $\bar{q}q^{(\rm subt)}$ the operator for which the power
divergences are subtracted.
The second and third terms represent a mixing with the identity
operator; 
the mass dimension is compensated by $m_q/a^2$ and $m_q^3$,
respectively. 
These coefficients $c_1^{(N)}$ and $c_2^{(N)}$ can be obtained by
fitting the vacuum expectation value $\langle\bar{q}q^{(N)}\rangle$
as a function of the valence quark mass $m_q$ \cite{Noaki:2009xi}.
When the correlation with the Wilson-loop is considered as in
(\ref{eq:diffLocalChiral}), the contribution from the identity
operator with the divergent coefficient 
$c_1^{(N)}m_q/a^2+c_2^{(N)}m_q^3$
cancels on the right hand side and the measurement is free from the
power divergences.

\begin{figure}
  \centering
  \includegraphics[width=0.49\textwidth,clip]{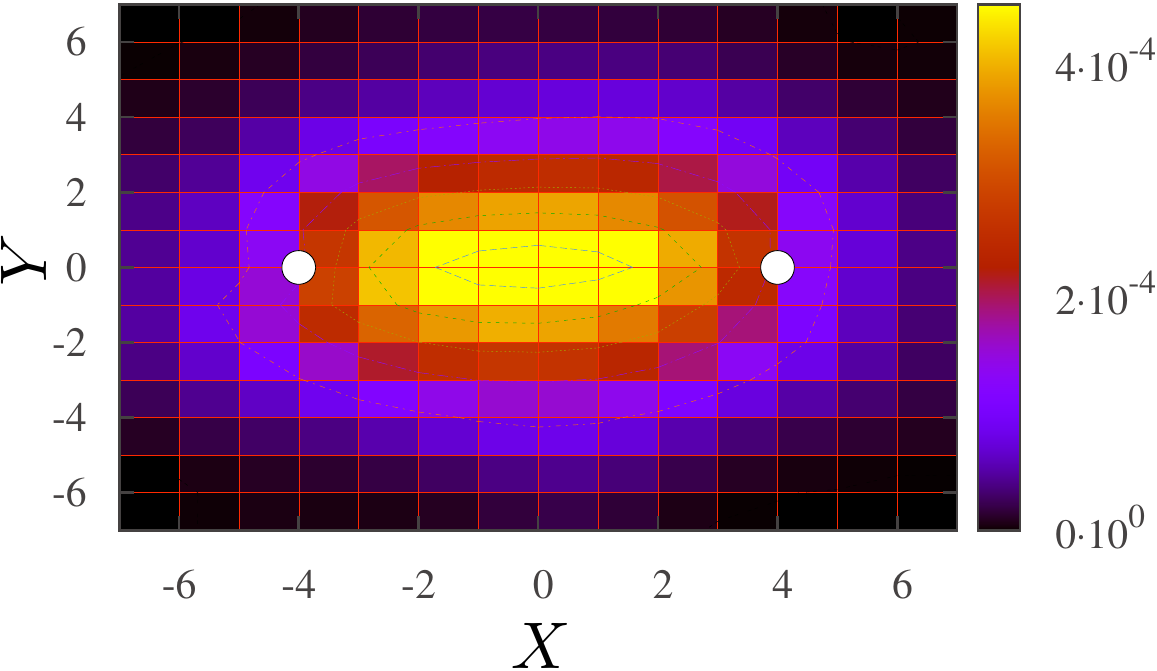}
  \caption{
    \label{fig:localChiral}
    Spatial profile of the local chiral condensate 
    $\langle\bar{q}q^{(N)}(\vec{x})\rangle_W$
    around a Wilson loop $W(R,T)$ with $R=8$ and $T=4$.
    The position of color sources are at $(X,Y) = (4,0)$ and $(-4,0)$,
    which are shown in the plot by white circles.
  }
\end{figure}

Figure \ref{fig:localChiral} shows 
the spatial distribution of $\langle\bar{q}q^{(N)}(\vec{x})\rangle_W$ 
on the $XY$-plane with a separation $R = 8$.
The location of color sources are shown by the circles.
In order to improve the signal of the Wilson loop, 
we apply the APE smearing for the spatial link-variables, 
and the temporal extent is fixed at $T=4$ for which the ground
state becomes dominant.
In this plot, $\bar{q}q(\vec{x})$ is set at $t = 0$, and the valence
quark mass is $m_q = 0.015a^{-1}$.

In order to improve the signal, the lattice data are averaged over
space-time.
Namely, assuming the translational invariance of the expectation
value, we shift the whole system including the Wilson-loop and the
local chiral condensate and take an average.
This can be done without additional computational cost to solve quark
propagators by using the low-lying eigenmodes.
This is one of the advantages of the construction (\ref{eq:localChiralCondensate}).

As Figure~\ref{fig:localChiral} demonstrates,
there appears a tube-like structure between the color sources,
where the change of the condensate becomes positive,
{\it i.e.} $\langle \bar{q}q^{(N)}(\vec{x}) \rangle_W > 0$.
It means that the magnitude of the chiral condensate is reduced
between the color charges, since $\langle\bar{q}q\rangle$ is negative in the vacuum.

Peak structures at the position of the charges are shown in the flux-tube 
measurements \cite{Bali:1994de,Haymaker:1994fm}
due to the strong enhancement of the action/energy density around the color charges.
In terms of the low-mode truncated local chiral condensate shown in Fig.~\ref{fig:localChiral}
no such characteristic structures around the color charges can be observed.
The absence of peaks will be discussed later.

The remaining logarithmic divergence in $\bar{q}q^{(\rm subt)}$ can be 
canceled by taking a ratio 
\begin{equation}
  r(\vec{x}) \equiv 
  \frac{
    \langle\bar{q}q^{(\rm subt)}(\vec{x})\rangle_W}{
    \langle\bar{q}q^{(\rm subt)}\rangle
  }
  =
  \frac{
    \langle\bar{q}q^{\rm (subt)}(\vec{x})W(R,T)\rangle}{
    \langle\bar{q}q^{\rm (subt)}\rangle \langle W(R,T)\rangle
  },
  \label{eq:reductionRatio}
\end{equation}
where $\langle\bar{q}q^{\rm (subt)}\rangle$ is obtained by fitting the
vacuum expectation value $\langle\bar{q}q\rangle$ 
to (\ref{eq:subtractChiral}) as a function of the valence quark mass
$m_q$.
As there are no remaining ultraviolet divergences, the ratio
$r(\vec{x})$ has a proper continuum limit.
Hereafter, we mainly use this quantity to quantitatively estimate the
restoration of chiral symmetry.

\begin{figure}
  \subfigure[\ cross section at $Y=0$]{
    \includegraphics[width=0.48\textwidth,clip]{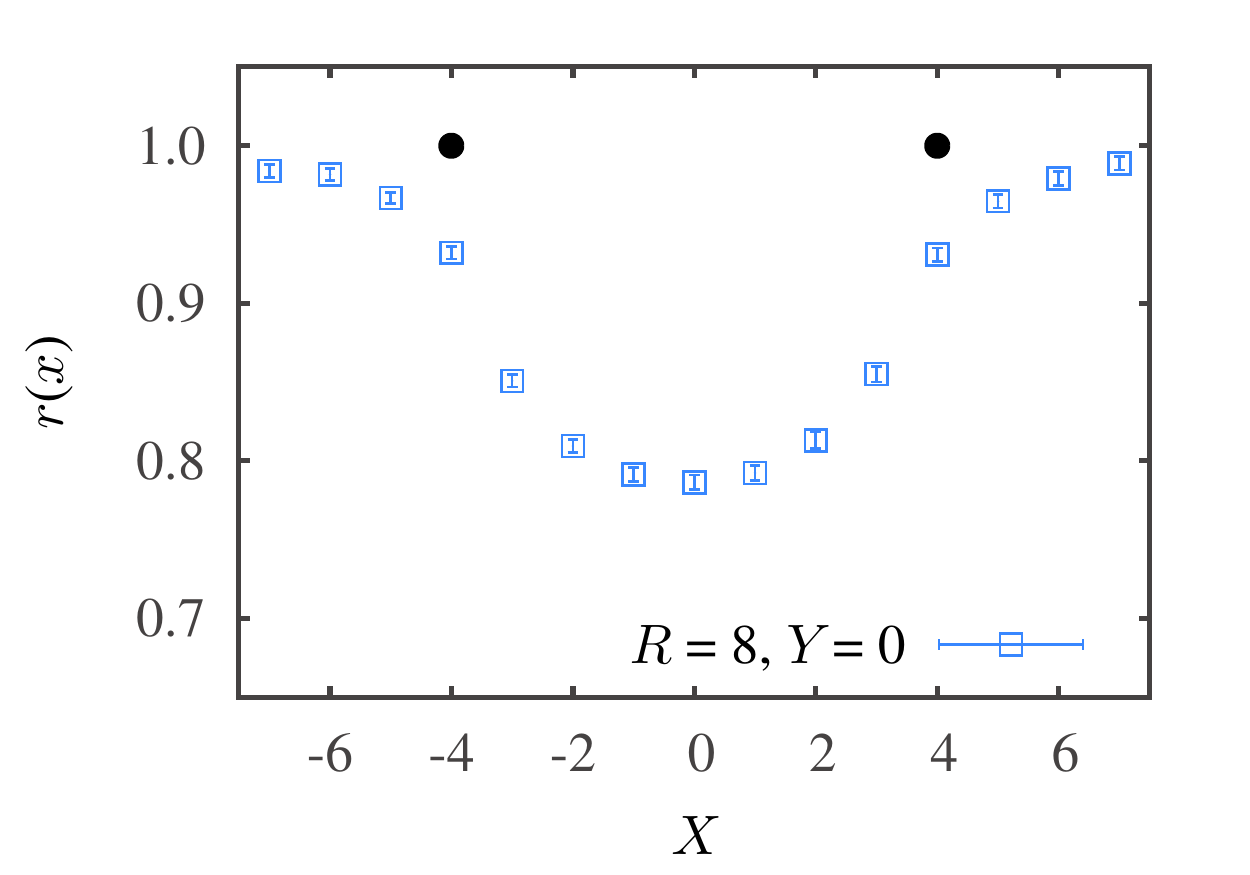}
  }
 \subfigure[\ cross section at $X=0$]{ 
     \includegraphics[width=0.48\textwidth,clip]{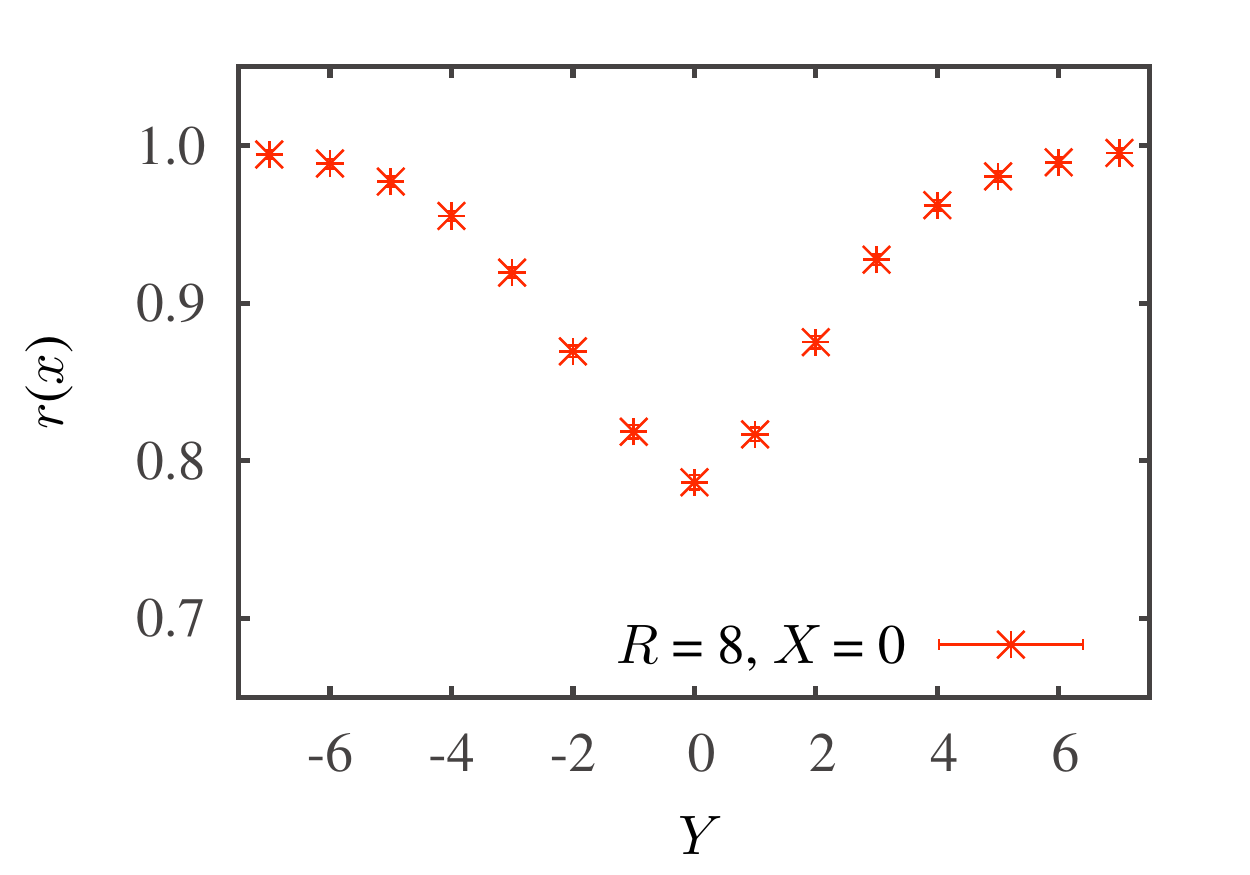}
}
  \caption{
    \label{fig:chiral_ratio}
    Ratio of the chiral condensate $r(x)$ for a separation $R = 8$.
    The color sources are separated along the $X$ axis and set at 
    $(X,Y) = (4,0)$ and $(-4,0)$.
    The plots show the cross section 
    (a) along the $X$-axis at $Y=0$ and (b) along $Y$-axis at $X=0$.
  }
\end{figure}

Figure~\ref{fig:chiral_ratio} shows the ratio $r(\vec{x})$ for 
the separation between the color sources fixed at $R=8$.
The plots 
Fig.~\ref{fig:chiral_ratio}~(a) and
Fig.~\ref{fig:chiral_ratio}~(b) 
correspond to the cross-sections of Figure~\ref{fig:localChiral}
along the $X$-axis and the transverse $Y$-axis.
The location of color sources is shown by black dots in
Fig.~\ref{fig:chiral_ratio}~(a).

These plots provide a quantitative measure of the reduction of the
chiral condensate.
The region where the chiral condensate is reduced forms a structure
that resembles the color flux-tube.
In other words, chiral symmetry is partially restored inside the
flux-tube. 
The restoration becomes stronger around the center of the flux,
which is about 20\% when $R=8$.

\begin{figure}
  \centering
  \subfigure[\ cross section at $Y=0$]{
    \includegraphics[width=0.48\textwidth,clip]{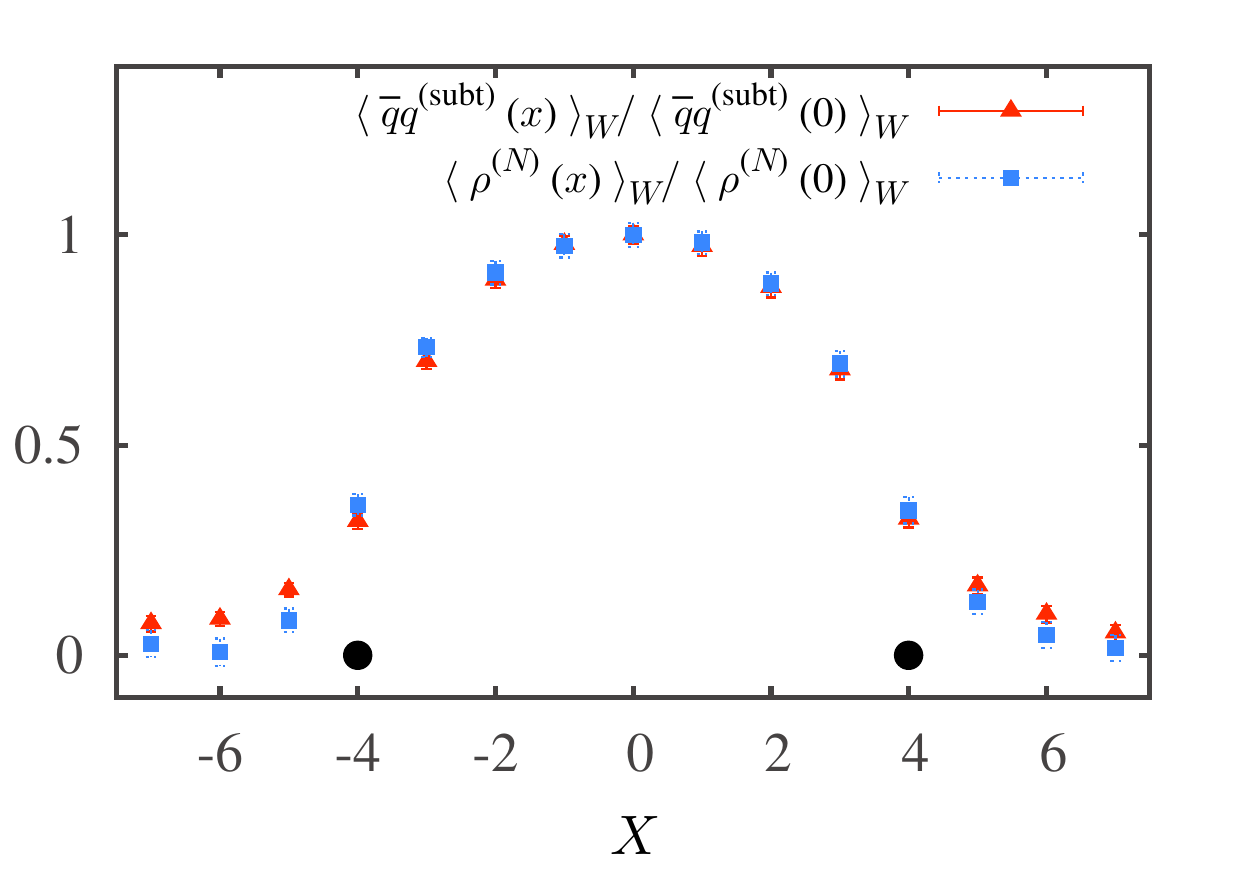}}
  \subfigure[\ cross section at $X=0$]{
    \includegraphics[width=0.48\textwidth,clip]{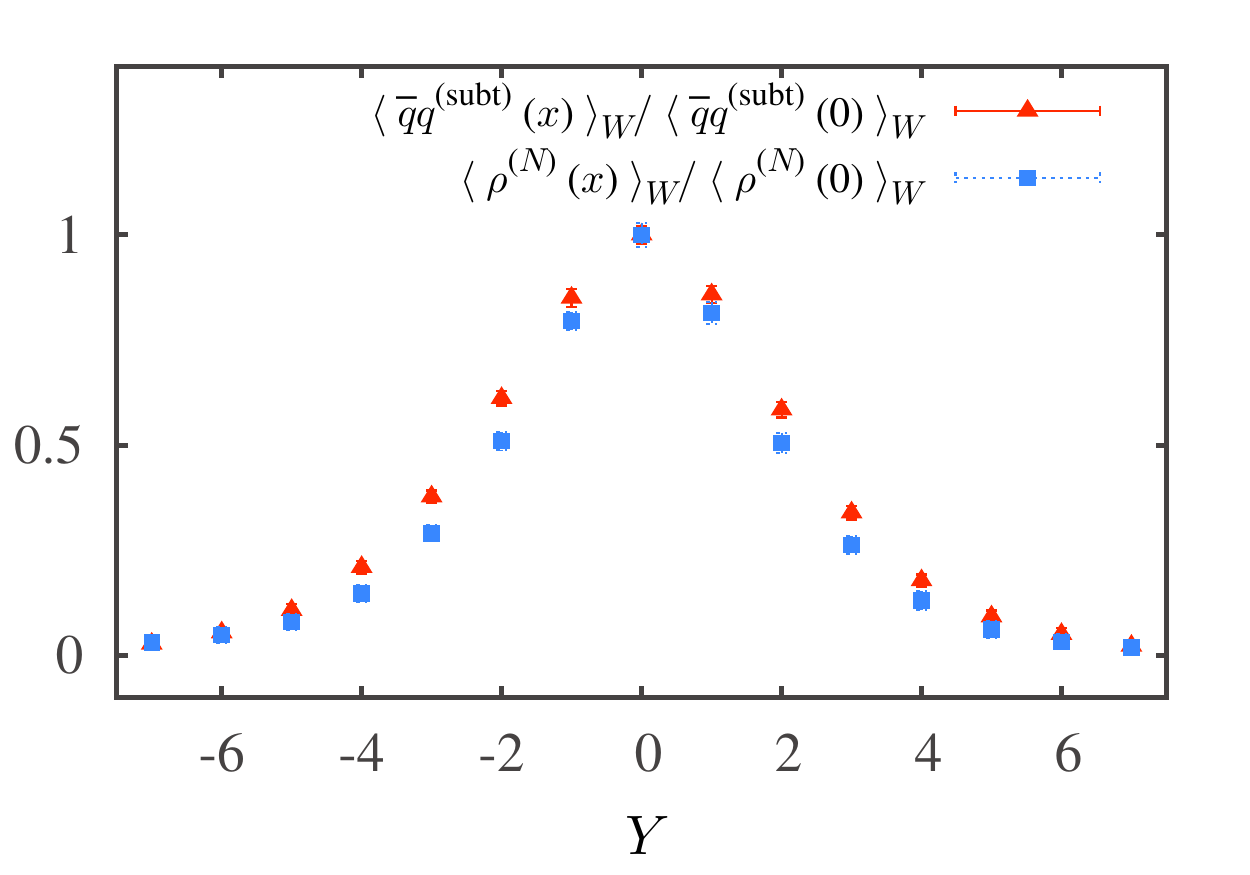}}
  \caption{
    \label{fig:chiral_action}
      The spatial profile of both local chiral condensate $\langle \bar{q}q^{\mathrm{(subt)}}(\vec{x}) \rangle_W$
      and UV Dirac mode truncated action density $\langle \rho^{(N)}(\vec{x}) \rangle_W$ around the color sources
      with a separation $R = 8$ using 160 low-lying eigenmodes.
      For a comparison of their shape, both quantities are normalized at the origin.
  }
\end{figure}

The close relationship between the flux-tube is suggested in Fig.~\ref{fig:chiral_action}.
We compare the cross-section 
of both chiral condensate $\langle \bar{q}q^{(\mathrm{subt})}(\vec{x}) \rangle_W$
and the action density defined by (\ref{eq:action_dens}) with a cutoff on the mode number
$\langle \rho^{(N)} (\vec{x})\rangle_W$
around the Wilson loop
using low-lying 160 eigenmodes.
The latter is calculated inserting
the action density $\rho(\vec{x})$
in place of $\bar{q}q(\vec{x})$ in Eq.~(\ref{eq:diffLocalChiral}),
which is used for the flux-tube measurement \cite{Bali:1994de,Haymaker:1994fm,Cea:1995zt}.
In order to compare the profile, both quantities are normalized to unity at the origin.
Apart from their normalization coefficients,
the spatial profile of the chiral condensate shows a good agreement with 
UV Dirac mode truncated action density.
As mentioned above, the action density is strongly enhanced around the color charges
as reported in \cite{Bali:1994de,Haymaker:1994fm}.
However, both UV filtered densities do not have such structures.
Our conclusion is that 
such peak mainly comes from ultraviolet divergent part and thus cannot be seen 
in Fig.~\ref{fig:chiral_action} (a) within our cutoff scale.

\begin{figure}[tb]
  \subfigure[\ cross section at $Y=0$]{
    \includegraphics[width=0.48\textwidth,clip]{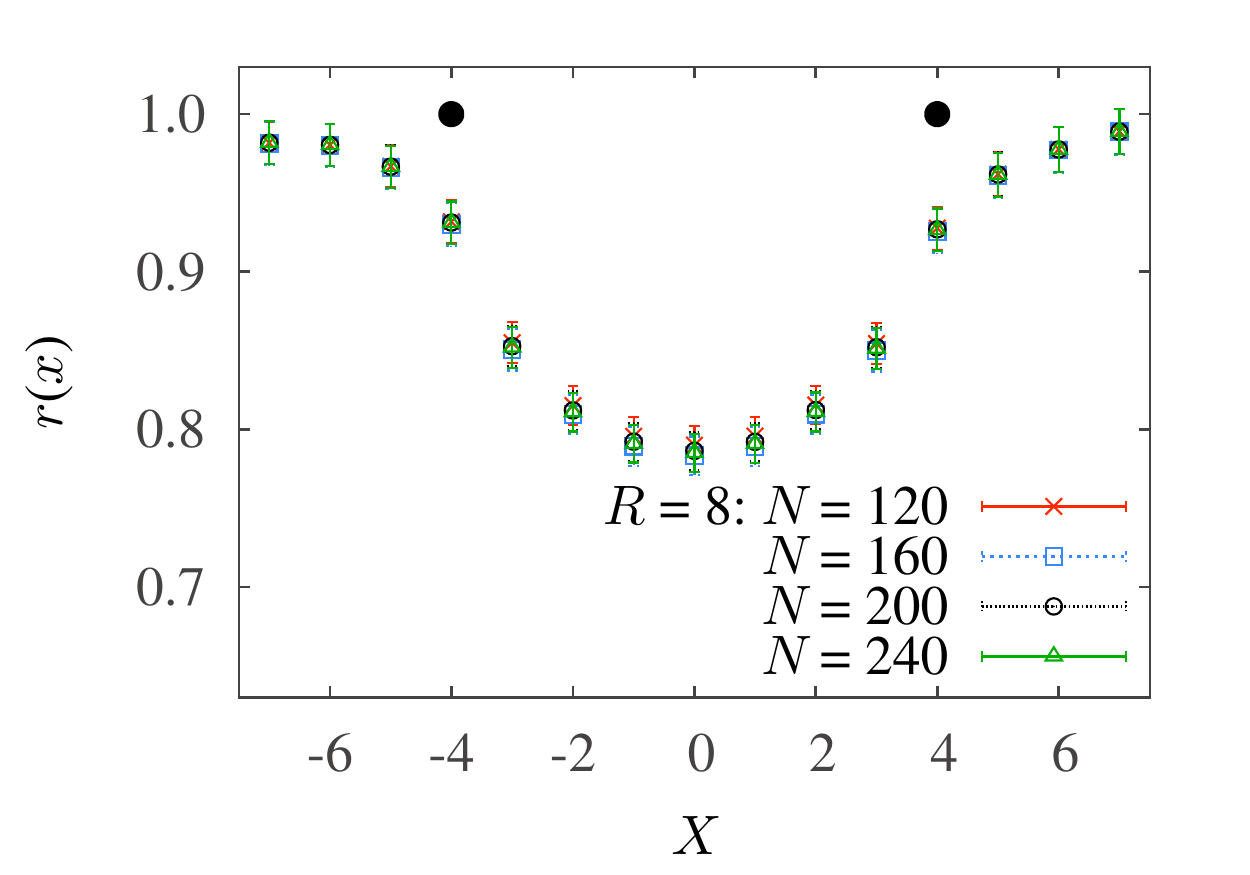}}
  \subfigure[\ cross section at $X=0$]{
    \includegraphics[width=0.48\textwidth,clip]{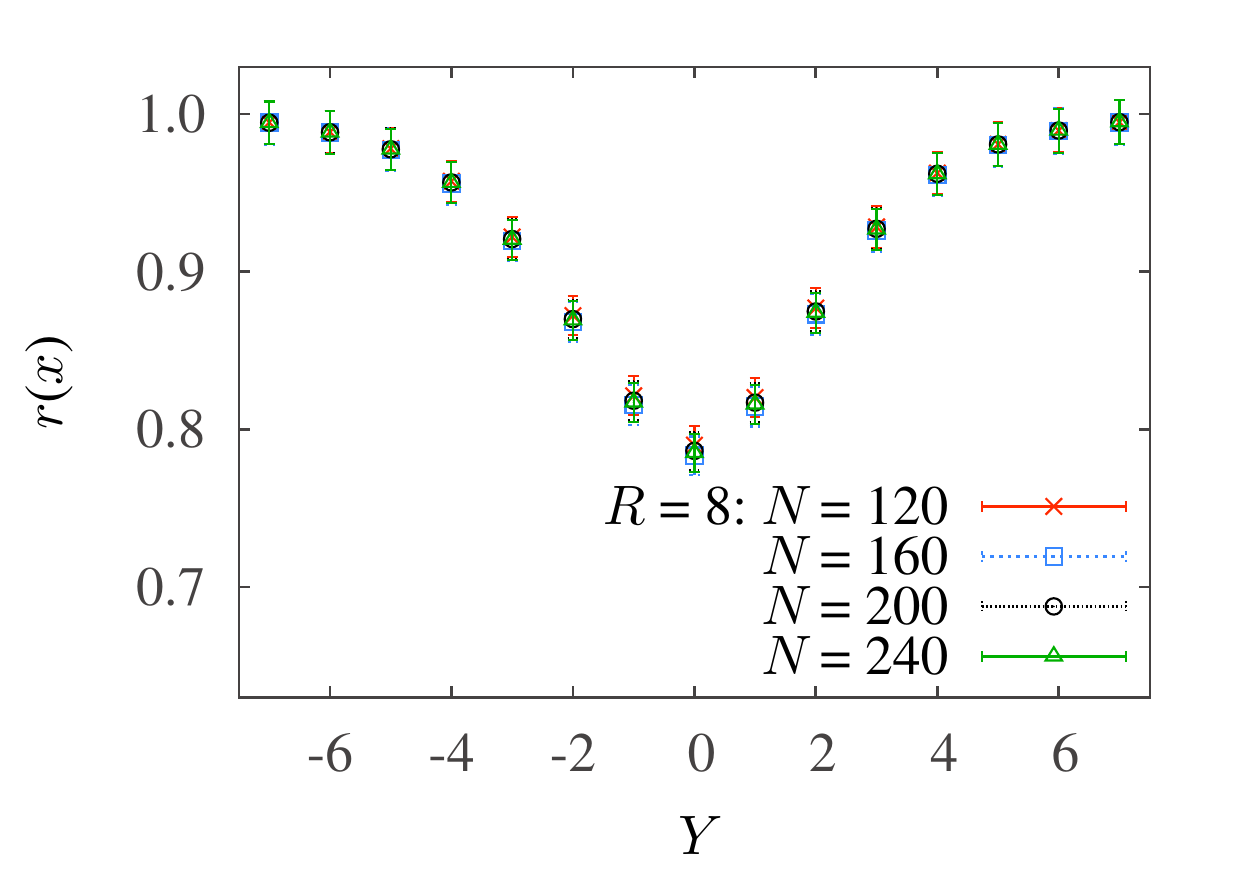}}
  \caption{
    \label{fig:chiral_ratio_cut}
    Same as Fig.~\ref{fig:chiral_ratio} but with different numbers
    of eigenmodes included:
    $N$ = 120, 160, 200, and 240.
  }
\end{figure}

Figure~\ref{fig:chiral_ratio_cut} shows the same plot as Fig.~\ref{fig:chiral_ratio}
but with different values of $N$, the number of eigenmodes included in the sum (\ref{eq:localChiralCondensate}).
As expected from the construction that cancels the ultraviolet
divergences, there is no significant difference between $N=120$ and 240.
Our choice $N=160$ is therefore sufficiently conservative
to estimate the local chiral condensate inside the tube.
Up to the largest eigenmode in our calculation at $N = 240$,
we have confirmed such saturation for other quantities considered in
this paper except for the magnitude of the action density $\rho^{(N)}(\vec{x})$ and topological charge density
$q_{\rm top}^{(N)}(\vec{x})$.
The value of these quantities strongly depends on the cutoff scale $\lambda_{\rm max}$
as expected from the definition in Eqs.~(\ref{eq:action_dens}) and (\ref{eq:topolo_dens}).
However, the spatial profile of both
$\langle \rho^{(N)}(\vec{x}) \rangle_W$ and $\langle \bar{q}q^{\mathrm{(subt)}}(\vec{x}) \rangle_W$
are rather stable,
and there are no signature of peaks within our truncation as in Fig.~\ref{fig:chiral_ratio_cut}~(a).

The partial restoration of chiral symmetry is in accordance with the
chiral bag model picture for the quark-antiquark system
\cite{Hosaka:1996ee}. 
In the na\"ive bag-model, chiral symmetry is completely restored
inside the bag, while
Fig.~\ref{fig:chiral_ratio} suggests a smooth boundary with a reduced
but non-zero condensate inside the bag.

\subsection{Chiral symmetry restoration as a function of the separation}
Next we study the chiral symmetry restoration depending on the
separation of the color sources.

\begin{figure}
  \includegraphics[width=0.47\textwidth,clip]{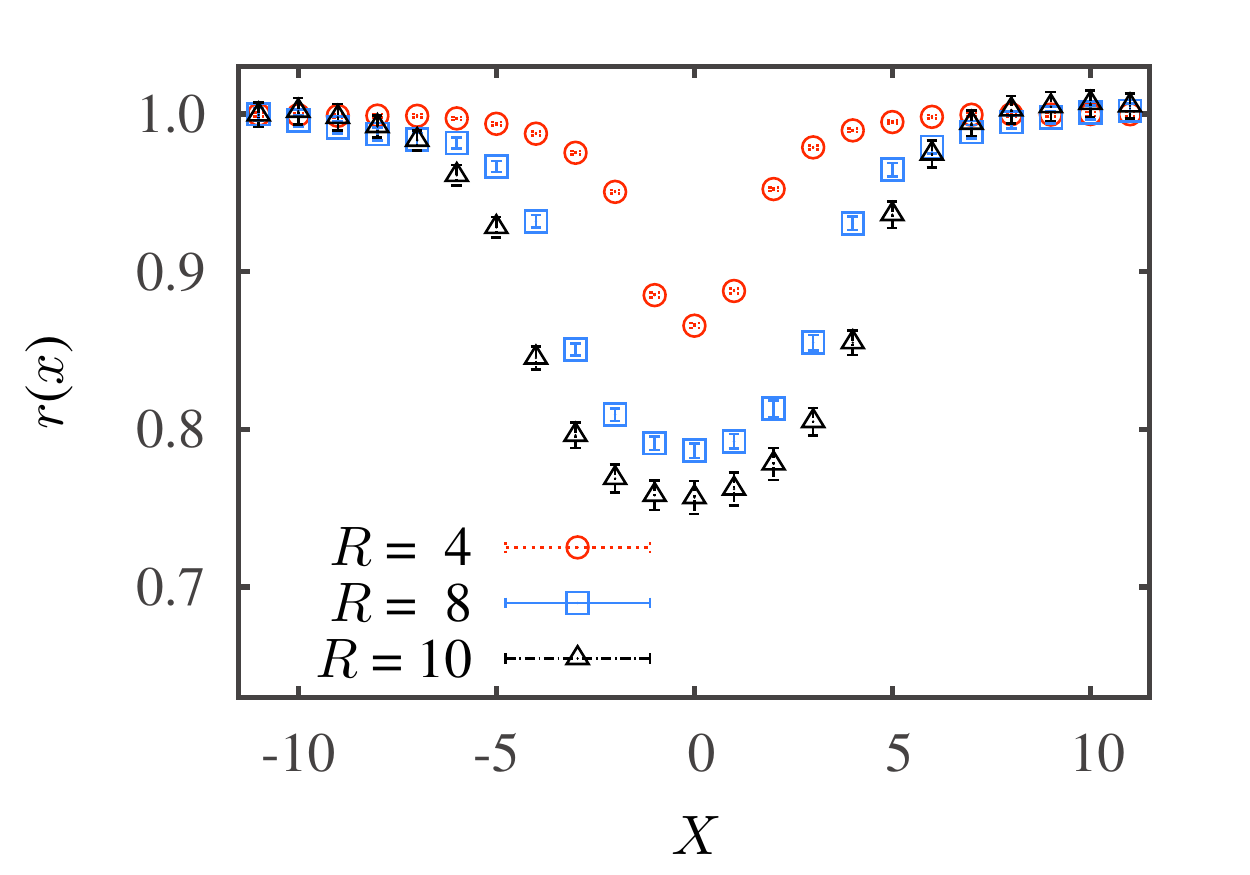}
  \caption{
    \label{fig:chiral_ratio_R_dep}
    Chiral condensate ratio $r(x)$ along the $X$-axis.
    The results with increasing separation $R$: $R$ = 4, 8 and 10.
    Color sources are located at $(R/2,0)$ and $(-R/2,0)$.
  }
\end{figure}

Figure \ref{fig:chiral_ratio_R_dep} compares the cross section of
$r(x)$ along the $X$-axis with $R$ = 4, 8 and 10.
By increasing the separation, we observe that the region of partial
restoration stretches between the color sources,
which are located at $X = R/2$ and $- R/2$.
This supports the picture of the tube structure.

The magnitude of the reduction increases with $R$.
For instance, at the origin, 
the reduction of about 15\% at $R = 4$ grows up to about 25\% 
at $R = 10$.
Beyond $R=10$, the statistical signal becomes much worse,
and the effect of spatial boundary would become important as $R$
approaches $L/2$.

\begin{figure}
  \centering
  \includegraphics[width=0.48\textwidth,clip]{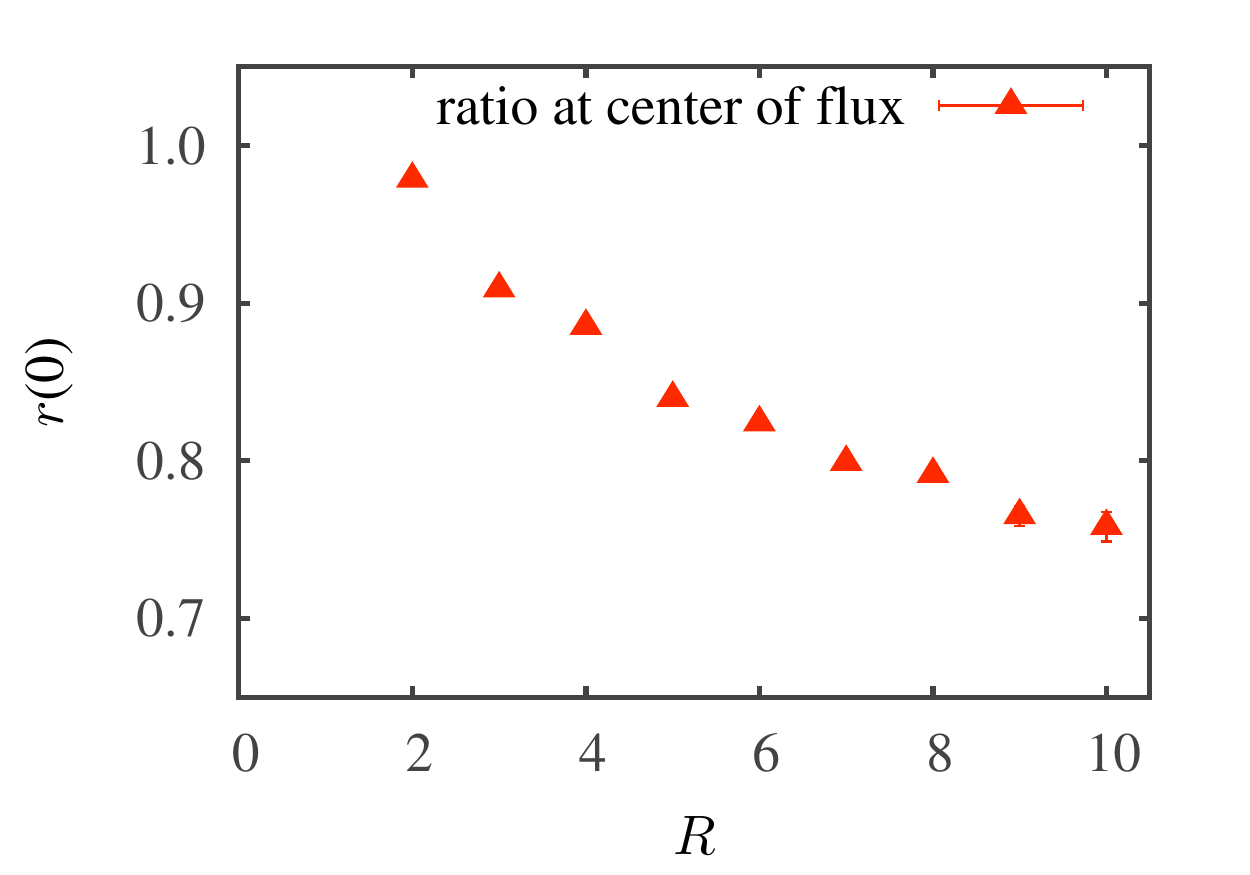}
  \caption{
    \label{fig:chiral_ratio_center}
    Ratio at the center of the flux $r(0)$
    as a function of the separation $R$.
  }
\end{figure}

In Figure~\ref{fig:chiral_ratio_center}, we plot
the value of the ratio at the center $r(0)$, where the magnitude becomes minimum, as a function of $R$.
As the separation $R$ increases, 
the ratio of the chiral condensate decreases monotonically until the
maximum distance we could explore.
At larger distances, the effect of string breaking should manifest
itself in dynamical QCD, and the local chiral condensate would stop
decreasing. 
As far as we can observe, the reduction of chiral condensate inside
the color flux-tube is of the size of 20--25\% at the distance of 1~fm,
assuming that the string breaking does not occur in this scale \cite{Bali:2005fu},
since it is difficult to observe the breaking state using the Wilson loop as a color source.

By increasing the separation between color sources, the
thickness of the flux is expected to grow logarithmically as a
function of its length \cite{Hasenfratz:1980ue,Luscher:1980iy}.
Such behavior has indeed been observed in quenched lattice QCD
calculations \cite{Cardoso:2013lla}
(see also \cite{Bakry:2010sp} for a study at finite temperature), 
for which one can increase the statistics more easily.
In this work we try to observe the thickness through the chiral
condensate ratio $r(x)$.

\begin{figure}
  \includegraphics[width=0.48\textwidth,clip]{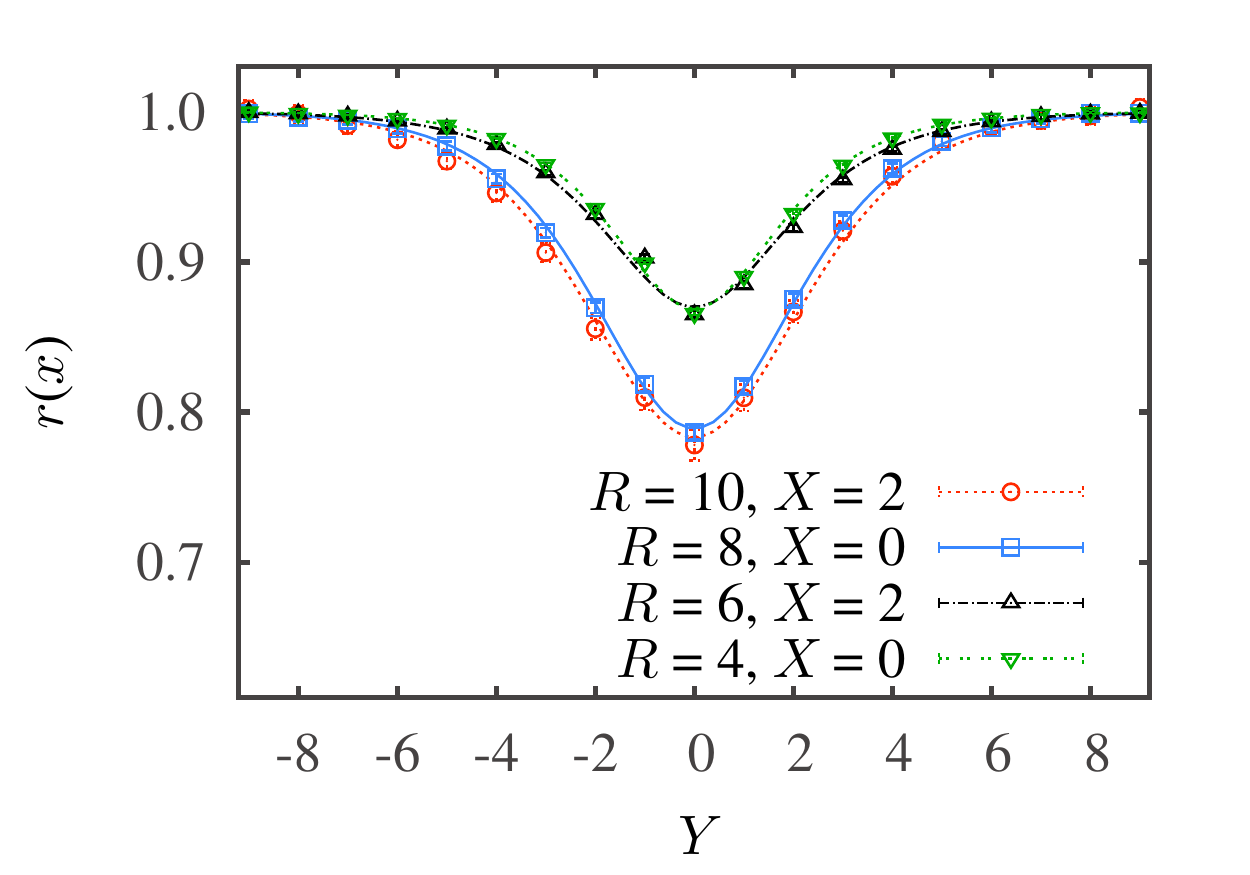}
  \caption{
    \label{fig:chiral_ratio_comp}
    Cross section of the ratio along the $Y$-axis with different
    separations $R$ and cut points $X$.
  }
\end{figure}

Figure~\ref{fig:chiral_ratio_comp} shows the cross-section of $r(x)$
along the $Y$-axis for some combinations of $R$ and $X$.
It is clear that for a fixed $X$ the flux is thicker when the
separation $R$ is larger.
More interestingly, the curve for $R = 8$ at $X = 0$ almost coincides
with that for $R = 10$ at $X = 2$.
Similarly, the curve for $R = 4$ at $X = 0$ coincides
with that for $R = 6$ at $X = 2$.
Due to the reflection symmetry, these behaviors are also observed at $X = - 2$.
This indicates that the thickness of the flux is highly correlated
with the magnitude of the reduction.
We also note that these corresponding cross-sections
have the same distance from the color charge, which is given by $R/2 - |X|$
for $|X| \leq R/2$.

In fact, such coincidence is expected from an effective string model
\cite{Cea:2012qw,Kharzeev:2014xta}.
According to that model, the ratio $r(Y)$ is written as
\begin{equation}
  r(Y) = 1 - \tilde{r} \frac{\mu^2}{\alpha}
  \frac{K_0((\mu^2 Y^2 + \alpha^2)^{1/2})}{K_1(\alpha)},
  \label{eq:fit_func}
\end{equation}
where $K_0(x)$ and $K_1(x)$ are modified Bessel functions.
The parameter $\mu$ has a physical interpretation as the inverse
penetration length of the flux from the perpendicular direction, and
$\alpha$ is the thickness of the core.
The parameter $\tilde{r}$ represents the strength of the condensate
reduction.

The function (\ref{eq:fit_func}) reproduces the lattice data quite
well, as shown by the curves in Fig.~\ref{fig:chiral_ratio_comp}.
The fit results are summarized in Table~\ref{tab:fit_cross_section}.
The penetration length is in the range of $1/\mu\simeq$
1.0--1.6, which corresponds to 0.11--0.18~fm in physical units.
The core size $\alpha$ is 1.4--3.6 in lattice units, and is in the
range 0.15--0.4~fm.
We observe an increase of $\alpha$ as $R$ increases while $X$ is fixed
at zero, but with the large statistical error we are not able to claim
the clear evidence of the string fattening.

\begin{table}[tbp]
  \begin{tabular}{ccc|cccc}
    \hline \hline
    $R$ & $X$ & $r(0)$ & $\tilde{r}$ & $\mu$ & $\alpha$ & $\chi^2/\mathrm{dof}$ \\
    \hline
    10 &0 &0.757(10) & 1.54(7) & 0.66(11) & 2.3(0.9) & 0.40 \\
    10 &1 &0.762(10) & 1.46(7) & 0.72(14) & 2.8(1.2) & 0.46 \\
    10 &2 &0.778(10) & 1.26(6) & 0.71(14) & 2.5(1.1) & 0.66 \\
    10 &3 &0.805(9)  & 1.00(6) & 0.75(18) & 2.5(1.3) & 1.00 \\
    \hline
    8 & 0 & 0.786(5) & 1.11(3) & 0.71(7) & 2.2(0.5) & 0.43 \\
    8 & 1 & 0.792(5) & 1.08(4) & 0.72(8) & 2.3(0.6) & 0.25 \\
    8 & 2 & 0.813(5) & 0.93(3) & 0.75(11) & 2.5(0.8) & 0.58 \\
    8 & 3 & 0.855(5) & 0.69(3) & 0.83(17) & 3.0(1.3) & 1.42 \\
    \hline
    6 & 0 & 0.815(3) & 0.89(2) & 0.66(4) & 1.7(2) & 0.38 \\
    6 & 1 & 0.827(3) & 0.81(2) & 0.65(4) & 1.6(2) & 1.01 \\
    6 & 2 & 0.865(3) & 0.65(2) & 0.61(4) & 1.4(2) & 2.36 \\
    \hline \hline
  \end{tabular}
  \caption{
    \label{tab:fit_cross_section}
    Fit results for $r(Y)$ to (\ref{eq:fit_func}) for each $R$ and
    $X$, together with $\chi^2$ per degrees of freedom.
    Note that $\mu$ and $\alpha$ are in lattice units.
    The condensate ratio at the center $r(0)$ is listed as well.
  }
\end{table}

\section{Chiral Condensate in the Three-Quark System}
\label{sec:4}

\subsection{Partial Restoration of chiral symmetry in the 3Q-system}
Next we consider a system consisting of three color charges that
represents a baryon system, that we call the 3Q-system in this paper.

Using the path-ordered product $U_k \equiv \prod_{\Gamma_k} e^{iagA_k}$
along a path $\Gamma_k$, the 3Q Wilson-loop is given by 
\begin{equation}
  W_\mathrm{3Q} \equiv 
  \frac{1}{3}\varepsilon_{abc}\varepsilon_{a'b'c'} U_1^{aa'} U_2^{bb'} U_3^{cc'},
  \label{}
\end{equation}
which is made color-singlet by the totally anti-symmetric tensor
$\varepsilon_{abc}$ of color indices $a$, $b$, and $c$ 
\cite{Takahashi:2000te,Takahashi:2002bw}.
Similar to the $\mathrm{\bar{Q}Q}$-system,
the spatial distribution of the chiral condensate for the 3Q-system is measured as
\begin{equation}
  \langle \bar{q}q(\vec{x}) \rangle_{{\rm 3Q}}
  \equiv 
  \frac{
    \langle \bar{q}q(\vec{x}) W_{\rm 3Q} \rangle
  }{
    \langle W_{\rm 3Q} \rangle
  }
  - \langle \bar{q}q \rangle,
  \label{eq:chiral_around_3Q}
\end{equation}
with the 3Q Wilson loop $W_{\rm 3Q}$.
The ratio of the chiral condensate in the 3Q-system
$r_\mathrm{3Q}(\vec{x})$,
for which the ultraviolet divergences cancel,
is then constructed by 
\begin{equation}
  r_\mathrm{3Q}({\vec{x}}) \equiv \frac{\langle \bar{q}q^{(\mathrm{subt})}(\vec{x})W_\mathrm{3Q} \rangle}{
    \langle \bar{q}q^\mathrm{(subt)}\rangle\langle W_\mathrm{3Q} \rangle}.
  \label{}
\end{equation}

\begin{figure}
  \includegraphics[width=0.47\textwidth,clip]{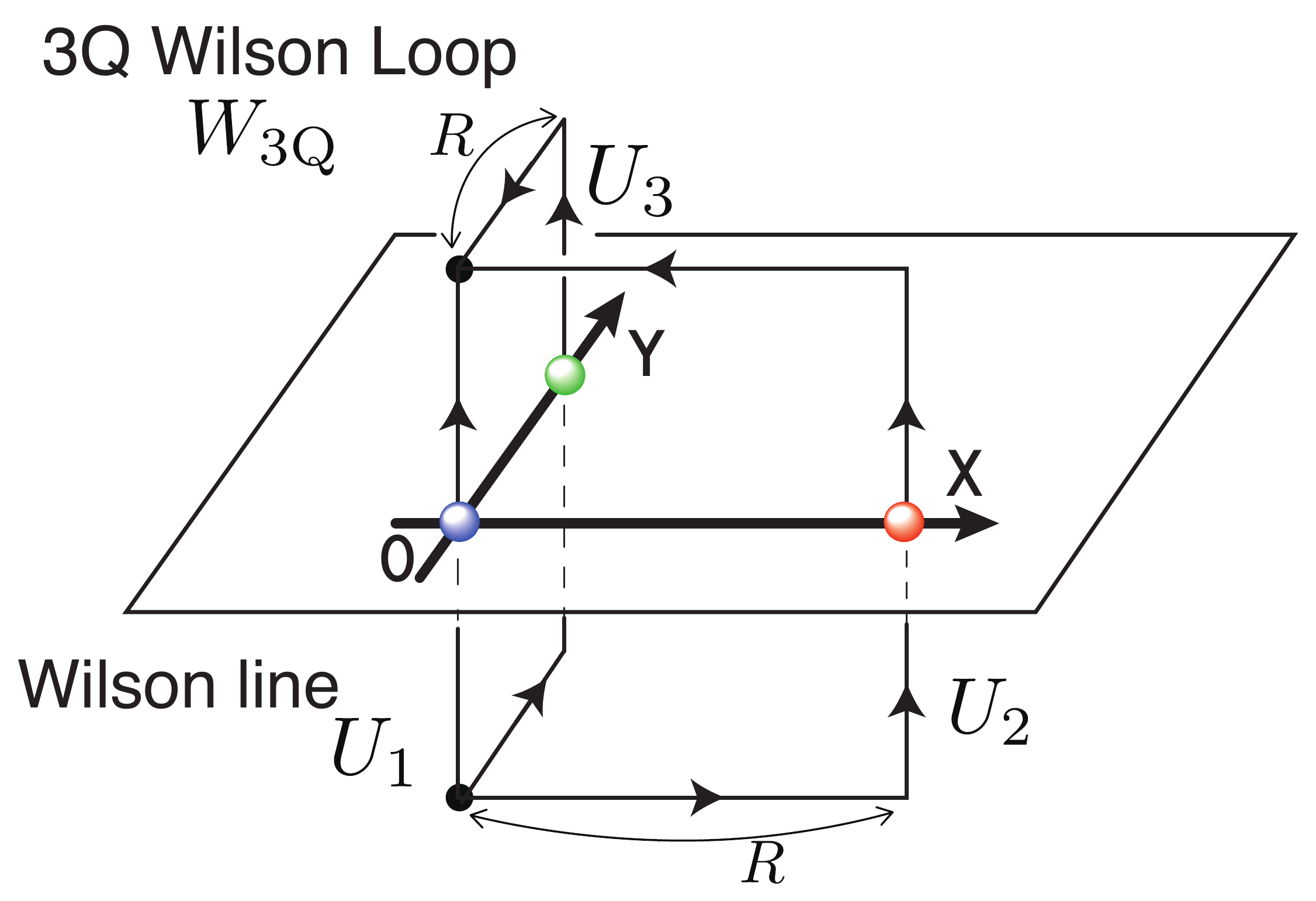}
  \caption{
    \label{fig:chiral_3q_measure}
    A schematic picture of the construction for three-quark system
    with an isosceles right triangle configuration.
    Three Wilson lines $U_k$ correspond to the static color sources.
  }
\end{figure}

Figure~\ref{fig:chiral_3q_measure} shows a schematic picture
of the construction of the 3Q Wilson loop $W_\mathrm{3Q}$ from the
Wilson lines $U_k$.
For simplicity, we use an isosceles right triangle configuration of
the color charges on the $XY$-plane,
and the coordinate is set as in Fig.~\ref{fig:chiral_3q_measure}.
In this case, the junction point of the three flux tubes (the Fermat point)
corresponds to the origin \cite{Takahashi:2000te,Takahashi:2002bw}.
The measurement of the local chiral condensate $\bar{q}q(\vec{x})$ 
is done at a fixed time slice.
The low-mode truncation number $N$, the temporal extension $T$ 
and other measurement setups are the same as in the
$\mathrm{Q\bar{Q}}$-system. 

\begin{figure}
  \includegraphics[width=0.48\textwidth,clip]{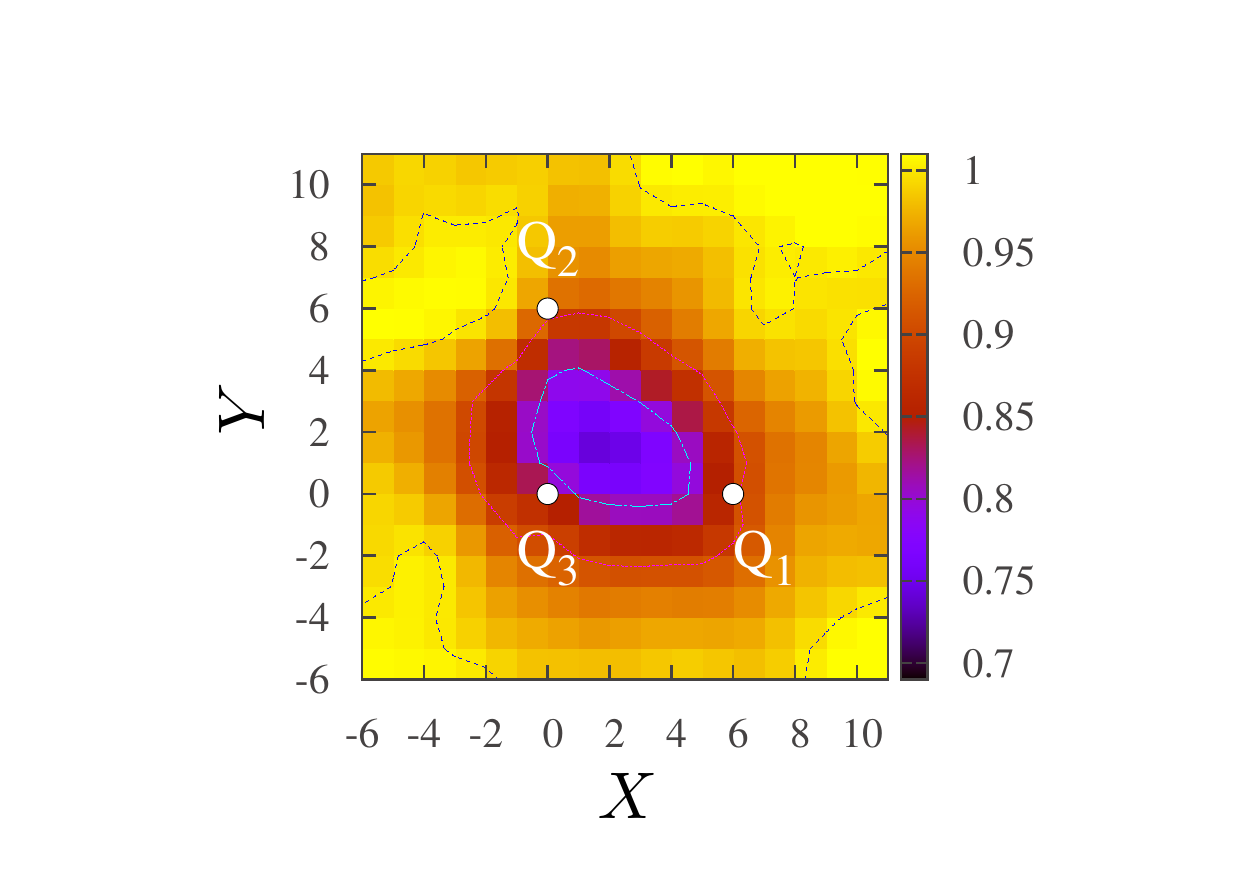}
  \caption{ \label{fig:qbarq_3Q_ratio}
  Condensate ratio $r_\mathrm{3Q}(\vec{x})$ with the color sources 
  at $Q_1 = (6,0)$, $Q_2 = (0,6)$, and $Q_3 = (0,0)$ on the $XY$-plane.
  }
\end{figure}

Figure \ref{fig:qbarq_3Q_ratio} shows the ratio 
$r_\mathrm{3Q}(\vec{x})$ with color sources 
at $(X,Y) = (6,0)$, $(0,6)$ and $(0,0)$ denoted by circles in the
plot.
As shown in Fig.~\ref{fig:qbarq_3Q_ratio},
the magnitude of the chiral condensate is reduced among the color
sources, which indicates the partial restoration of chiral
symmetry inside the 3Q-system.
Similar to the $\mathrm{Q\bar{Q}}$-system in Sec. III,
there appear no peaks at the color charges within our truncation scale.
We note that the characteristic $Y$-type flux is not clearly seen
in this plot, probably because the thickness of flux is comparable to
the color source separation.
Because of the statistical noise, we are not able to repeat the calculations increasing the quark separations.

Like in the $\mathrm{\bar{Q}Q}$-system, the magnitude of the
restoration depends on the separation of the sources.
Figure~\ref{fig:qbarq_3Q_ratio} shows the cross-section of the ratio
$r_\mathrm{3Q}(\vec{x})$ along the line of $X = Y$
with the color sources at $(X,Y) = (R,0)$, $(0,R)$ and $(0,0)$.
In this setup, the measurement goes through one color charge and the
center of mass of the system.
By comparing the data for $R = 3$ and 6, we find that the reduction is
more substantial for $R=6$, which is similar to the 
$\mathrm{Q\bar{Q}}$-system (see Fig.~\ref{fig:chiral_ratio_R_dep}).
The reduction of the local chiral condensate becomes larger with the
size of the loop, and take its minimum value at around the center of
gravity. 
With $R=6$, the reduction is about 30\%, which is also similar to that of
the $\mathrm{Q\bar{Q}}$-system.

\begin{figure}
  \includegraphics[width=0.48\textwidth,clip]{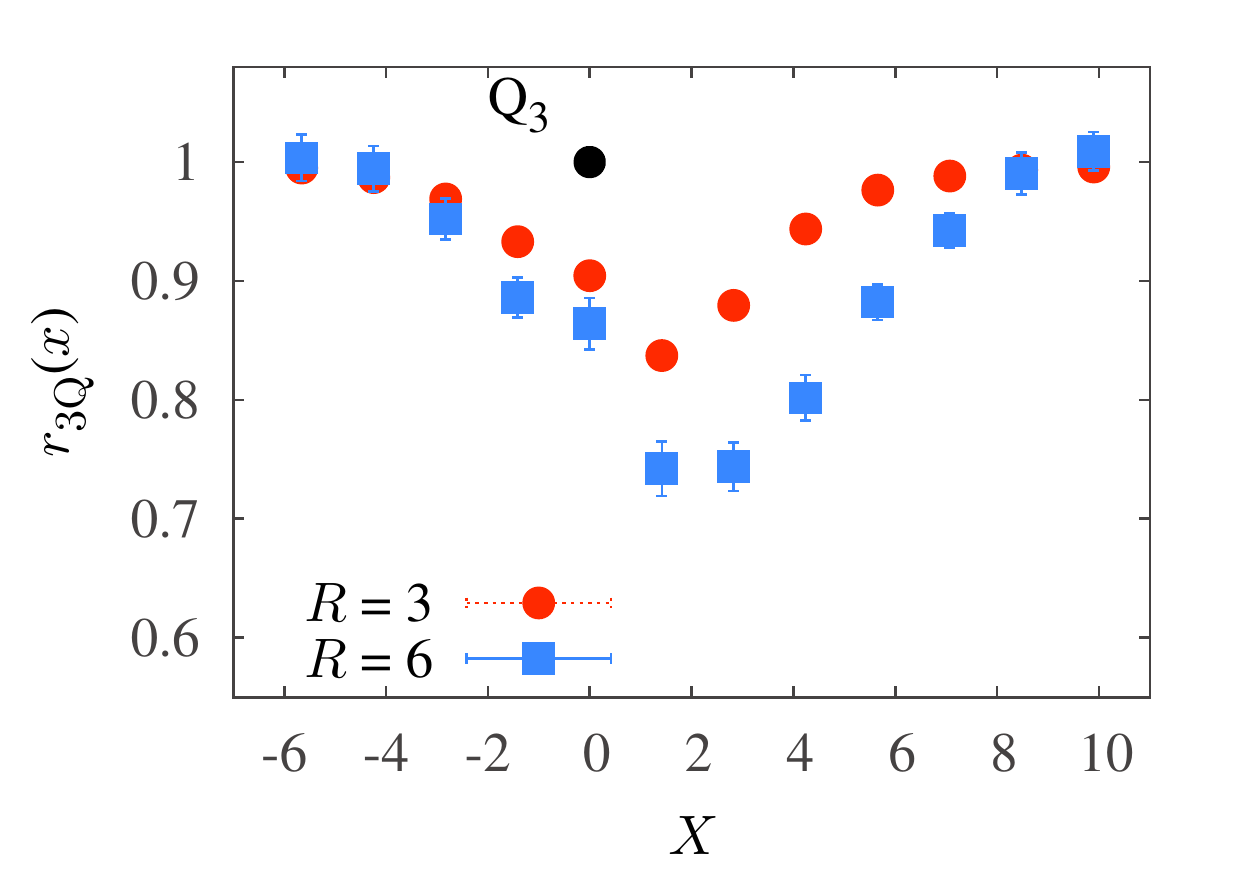}
  \caption{ \label{fig:qbarq_3Q_ratio_Rdep}
    Chiral condensate ratio $r_{\rm 3Q}(\vec{x})$ along 
    the line of $X = Y$ with the color sources at 
    $(R,0), (0,R)$ and $(0,0)$ on the $XY$-plane with $R = 3$ and 6.
  }
\end{figure}

\subsection{Partial restoration at \textit{finite density}}
Finally, using the observed modification of the local chiral
condensate around the color sources,
we estimate the size of the partial restoration of chiral symmetry in
\textit{finite density} QCD.
We consider the system of fixed number of baryons in a finite 
volume box, so that the baryon number density $\rho$ is $N_b/L^3$,
where $N_b$ is the number of baryons and $L^3$ is the spatial volume.
As a toy example we take $N_b=1$ and replace the baryon by the 3Q
Wilson-loop.
This only gives a crude approximation of the realistic system, but
given the difficulty of simulating QCD at finite chemical potential
it may provide a useful clue to the understanding of the finite
density QCD. 

The net change of the condensate under such system is estimated 
by the spatial average of the condensate ratio $r_\mathrm{3Q}(\vec{x})$:
\begin{eqnarray}
  \frac{\langle \bar{q}q \rangle_\rho}{\langle \bar{q}q \rangle_0}
  \equiv \frac{1}{L^3}\sum_{\vec{x}}^{L^3} r_{\mathrm{3Q}}(\vec{x}),
  \label{eq:spatial_average}
\end{eqnarray}
where $\langle\bar{q}q\rangle_\rho$ is the condensate at the finite
baryon number density $\rho=1/L^3$.
We use two lattice volumes, $L^3=16^3$ and $24^3$, which 
correspond to 
$(16a)^{-3} \simeq 0.18\ \mathrm{fm}^{-3}$
and
$(24a)^{-3} \simeq 0.05\ \mathrm{fm}^{-3}$, respectively.
The $16^3$ lattice roughly corresponds to the normal nuclear density  $\rho_0\simeq 0.18\ \mathrm{fm}^{-3}$.

\begin{figure}
  \includegraphics[width=0.47\textwidth,clip]{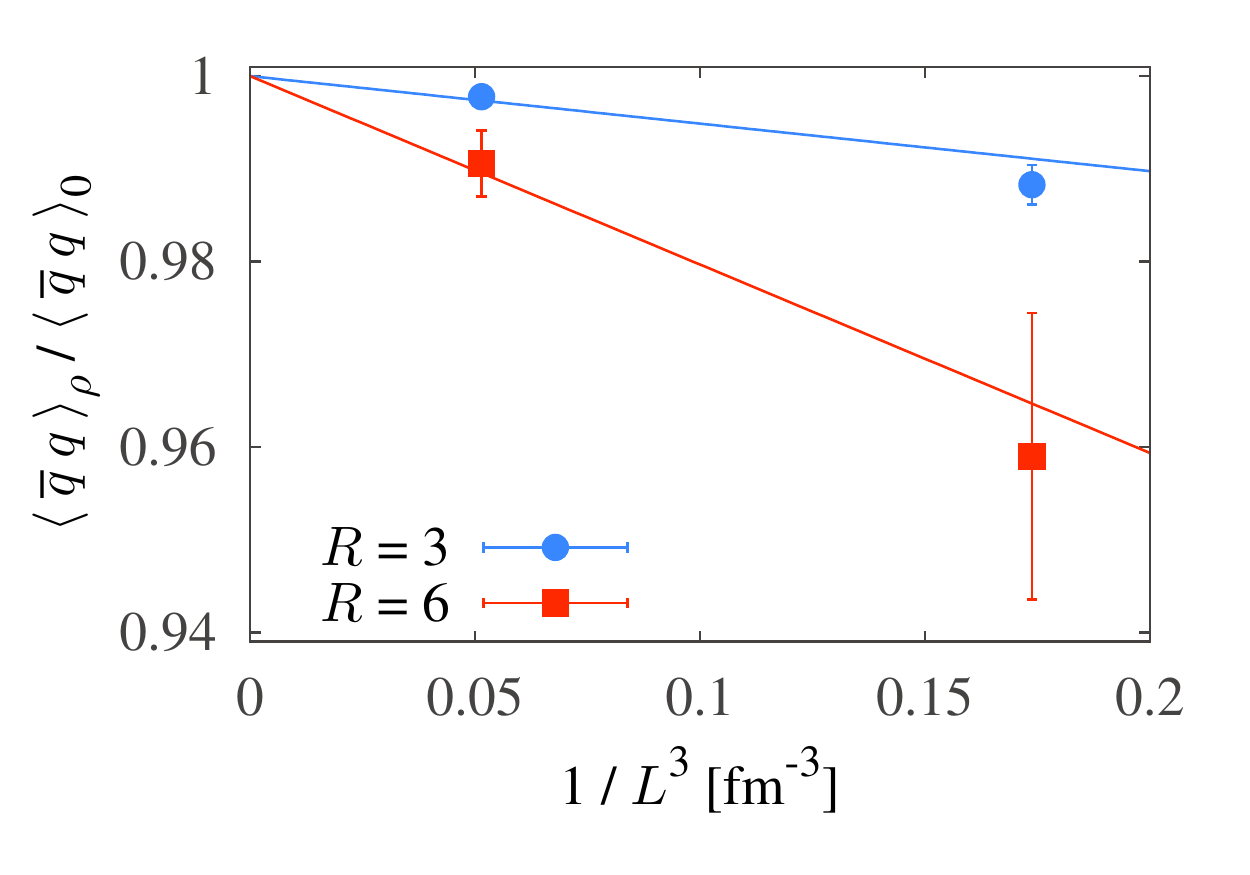}
  \caption{
    \label{fig:chiral_volume_dep}
    Reduction of the chiral condensate at finite density measured by
    $\langle \bar{q}q \rangle_\rho/\langle \bar{q}q \rangle_0$.
    The estimates for a given configuration of color sources, 
    {\it i.e.} at $(0,0), (R,0)$ and $(0,R)$ on the $XY$-plane,
    with $R=3$ and $6$ are shown.
  }
\end{figure}

Figure~\ref{fig:chiral_volume_dep} shows
$\langle \bar{q}q \rangle_\rho/\langle \bar{q}q \rangle_0$
as a function of $1/L^3$.
The two symbols correspond to the different configurations of the
color sources, {\it i.e.}
$(0,0)$, $(R,0)$ and $(0,R)$ with $R = 3$ and 6 on the $XY$-plane.
The solid lines are the results of a linear fit
with fixed value of 1 at $1/L^3 = 0$.
The linear dependence from unity at $\rho=0$ simply means that there
is a finite region where the chiral condensate is reduced from its
vacuum value.
Since the region gets larger with increasing $R$, the slope for larger
$R$ is steeper.

In our setup, the reduction of the chiral condensate at the normal
nuclear density is only $\sim$ 5\%, which is much smaller than the
phenomenological model estimate of the order of 30\%
\cite{Hayano:2008vn}.
Our estimate, however, assumes a fixed spatial size of the 
{\it baryon} which is smaller than the realistic nucleon.
For instance, the mean root square radius of our setup
in Fig.~\ref{fig:chiral_volume_dep} is 0.44~fm when $R=6$, while the
charge radius of proton is 0.88~fm.
As the restoration of chiral condensate is stronger for larger
separation, this suggests that $\langle\bar{q}q\rangle_\rho$ in 
realistic finite density QCD could be substantially lower than our
estimate. 

\section{Summary}
\label{sec:summary}
The Dirac eigenmodes carry the full information of the background
gauge field.
Indeed, having the complete set of the eigenvalue and eigenvectors,
one can reconstruct the field-strength tensor $F_{\mu\nu}(x)$ at any
point $x$.
They therefore offer an interesting way of filtering out the
ultraviolet modes and investigating the low-energy dynamics of QCD
by only using the low-lying eigenmodes upon reconstruction.
This is a sound and well-defined regularization method of quantum
field theory.

We use this method to investigate the spatial profile of the chiral
condensate under the presence of external sources.
On the lattices generated with 2+1 flavors of dynamical overlap
fermions, we calculate the low-lying eigenvalues and associated
eigenvectors of the overlap-Dirac operator, and use them to
reconstruct the chiral condensate locally.
Then, it is straightforward to measure its correlation with the
external color sources set up to model the $\mathrm{\bar{Q}Q}$ and 3Q
systems.

We find that the local chiral condensate shows a structure interpreted
as a color flux-tube between the $\mathrm{\bar{Q}Q}$ color sources, 
in which the condensate decreases significantly.
It indicates a partial restoration of chiral symmetry inside the flux-tube
and suggests that it happens also inside hadrons.
The spatial profile is consistent with a string model of the
confinement potential, giving another support for the presence of the
color flux-tube.

We perform a similar measurement in the 3Q system, which is new as far
as we have noticed.
It again shows the partial restoration of chiral symmetry among the
color sources.
The reduction of condensate is about 30\% for the separation between
the color sources of $\sim$ 1~fm.
It can be used to estimate the chiral condensate in the finite density
system.

The method developed in this work may easily be applied for the study
of finite temperature QCD, where Polyakov loops can be used for a
static color source.
Since the eigenmodes can be applied to define various charges,
such as the axial charge density, the quark number density,
and the topological charge, 
it may provide an interesting alternative to measure their spatial
distribution.

\section*{Acknowledgements}
The lattice QCD calculations have been done on SR16000 at 
High Energy Accelerator Research Organization (KEK)
under a support of its Large Scale Simulation Program (No. 13/14-05).
This work is supported in part by the Grant-in-Aid of the Japanese
Ministry of Education (No. 25287046 and 26247043),
and the SPIRE (Strategic Program for Innovative REsearch) Field 5
project.


\begin{thebibliography}{99}
\bibitem{Banks:1979yr} 
  T.~Banks and A.~Casher,
  Nucl.\ Phys.\ B {\bf 169}, 103 (1980).


\bibitem{Fukushima:2010bq} 
  K.~Fukushima and T.~Hatsuda,
  Rept.\ Prog.\ Phys.\  {\bf 74}, 014001 (2011)
  [arXiv:1005.4814 [hep-ph]].


\bibitem{Bali:1994de} 
  G.~S.~Bali, K.~Schilling and C.~Schlichter,
  Phys.\ Rev.\ D {\bf 51}, 5165 (1995)
  [hep-lat/9409005].

\bibitem{Haymaker:1994fm} 
  R.~W.~Haymaker, V.~Singh, Y.~C.~Peng and J.~Wosiek,
  Phys.\ Rev.\ D {\bf 53}, 389 (1996)
  [hep-lat/9406021].

\bibitem{Cea:1995zt} 
  P.~Cea and L.~Cosmai,
  Phys.\ Rev.\ D {\bf 52}, 5152 (1995)
  [hep-lat/9504008].

\bibitem{Gattringer:2002gn} 
  C.~Gattringer,
  Phys.\ Rev.\ Lett.\  {\bf 88}, 221601 (2002)
  [hep-lat/0202002].


\bibitem{Aoki:2012pma} 
  S.~Aoki, T.~W.~Chiu, G.~Cossu, X.~Feng, H.~Fukaya, S.~Hashimoto, T.~H.~Hsieh and T.~Kaneko {\it et al.},
  PTEP {\bf 2012}, 01A106 (2012).

\bibitem{Neuberger:1997fp} 
  H.~Neuberger,
  Phys.\ Lett.\ B {\bf 417}, 141 (1998)
  [hep-lat/9707022].

\bibitem{Neuberger:1998wv} 
  H.~Neuberger,
  Phys.\ Lett.\ B {\bf 427}, 353 (1998)
  [hep-lat/9801031].



\bibitem{Fukaya:2007fb} 
  H.~Fukaya {\it et al.}  [JLQCD Collaboration],
  Phys.\ Rev.\ Lett.\  {\bf 98}, 172001 (2007)
  [hep-lat/0702003].

\bibitem{Fukaya:2007yv} 
  H.~Fukaya {\it et al.}  [TWQCD Collaboration],
  Phys.\ Rev.\ D {\bf 76}, 054503 (2007)
  [arXiv:0705.3322 [hep-lat]].

\bibitem{Fukaya:2009fh} 
  H.~Fukaya {\it et al.}  [JLQCD Collaboration],
  Phys.\ Rev.\ Lett.\  {\bf 104}, 122002 (2010)
  [Erratum-ibid.\  {\bf 105}, 159901 (2010)]
  [arXiv:0911.5555 [hep-lat]].

\bibitem{Fukaya:2010na} 
  H.~Fukaya {\it et al.}  [JLQCD and TWQCD Collaborations],
  Phys.\ Rev.\ D {\bf 83}, 074501 (2011)
  [arXiv:1012.4052 [hep-lat]].

\bibitem{Iritani:2013rla} 
  T.~Iritani, G.~Cossu and S.~Hashimoto,
  PoS LATTICE {\bf 2013}, 376 (2014)
  [arXiv:1311.0218 [hep-lat]].

\bibitem{Iritani:2014jqa} 
  T.~Iritani, G.~Cossu and S.~Hashimoto,
  PoS Hadron {\bf 2013}, 159 (2013)
  [arXiv:1401.4293 [hep-lat]].

\bibitem{Iritani:2014fga} 
  T.~Iritani, G.~Cossu and S.~Hashimoto,
  PoS (Lattice 2014) {\bf 338} [arXiv:1412.2322 [hep-lat]].

\bibitem{Niedermayer:1998bi} 
  F.~Niedermayer,
  Nucl.\ Phys.\ Proc.\ Suppl.\  {\bf 73}, 105 (1999)
  [hep-lat/9810026].


\bibitem{Fukaya:2006vs} 
  H.~Fukaya {\it et al.}  [JLQCD Collaboration],
  Phys.\ Rev.\ D {\bf 74}, 094505 (2006)
  [hep-lat/0607020].

\bibitem{Aoki:2007ka} 
  S.~Aoki, H.~Fukaya, S.~Hashimoto and T.~Onogi,
  Phys.\ Rev.\ D {\bf 76}, 054508 (2007)
  [arXiv:0707.0396 [hep-lat]].

\bibitem{Ilgenfritz:2007xu} 
  E.-M.~Ilgenfritz, K.~Koller, Y.~Koma, G.~Schierholz, T.~Streuer and V.~Weinberg,
  Phys.\ Rev.\ D {\bf 76}, 034506 (2007)
  [arXiv:0705.0018 [hep-lat]].

\bibitem{Ilgenfritz:2008ia} 
  E.-M.~Ilgenfritz, D.~Leinweber, P.~Moran, K.~Koller, G.~Schierholz and V.~Weinberg,
  Phys.\ Rev.\ D {\bf 77}, 074502 (2008)
  [Erratum-ibid.\ D {\bf 77}, 099902 (2008)]
  [arXiv:0801.1725 [hep-lat]].

\bibitem{Feilmair:1988js} 
  W.~Feilmair, M.~Faber and H.~Markum,
  Phys.\ Rev.\ D {\bf 39}, 1409 (1989).

\bibitem{Sakuler:1992qx} 
  W.~Sakuler, W.~Burger, M.~Faber, H.~Markum, M.~M\"uller, P.~De Forcrand, A.~Nakamura and I.~O.~Stamatescu,
  Phys.\ Lett.\ B {\bf 276}, 155 (1992).

\bibitem{Faber:1993sw} 
  M.~Faber, M.~Schaler and H.~Gausterer,
  Phys.\ Lett.\ B {\bf 317}, 409 (1993).

\bibitem{Chagdaa:2006zz} 
  S.~Chagdaa and E.~Laermann,
  PoS LAT {\bf 2007}, 172 (2007).

\bibitem{Noaki:2009xi} 
  J.~Noaki, T.~W.~Chiu, H.~Fukaya, S.~Hashimoto, H.~Matsufuru, T.~Onogi, E.~Shintani and N.~Yamada,
  Phys.\ Rev.\ D {\bf 81}, 034502 (2010)
  [arXiv:0907.2751 [hep-lat]].

\bibitem{Hosaka:1996ee} 
  A.~Hosaka and H.~Toki,
  Phys.\ Rept.\  {\bf 277}, 65 (1996).

\bibitem{Bali:2005fu} 
  G.~S.~Bali {\it et al.}  [SESAM Collaboration],
  Phys.\ Rev.\ D {\bf 71}, 114513 (2005) [hep-lat/0505012].



\bibitem{Hasenfratz:1980ue}
  A.~Hasenfratz, E.~Hasenfratz and P.~Hasenfratz,
  Nucl.\ Phys.\ B {\bf 180} (1981) 353.

\bibitem{Luscher:1980iy}
  M.~L\"uscher, G.~M\"unster and P.~Weisz,
  Nucl. Phys.\ B {\bf 180} (1981) 1.

\bibitem{Cardoso:2013lla} 
  N.~Cardoso, M.~Cardoso and P.~Bicudo,
  Phys.\ Rev.\ D {\bf 88}, 054504 (2013)
  [arXiv:1302.3633 [hep-lat]].

\bibitem{Bakry:2010sp} 
  A.~S.~Bakry, D.~B.~Leinweber and A.~G.~Williams,
  Phys.\ Rev.\ D {\bf 85}, 034504 (2012)
  [arXiv:1011.1380 [hep-lat]].

\bibitem{Cea:2012qw} 
  P.~Cea, L.~Cosmai and A.~Papa,
  Phys.\ Rev.\ D {\bf 86}, 054501 (2012)
  [arXiv:1208.1362 [hep-lat]].

\bibitem{Kharzeev:2014xta} 
  D.~E.~Kharzeev and F.~Loshaj,
  Phys.\ Rev.\ D {\bf 90}, 037501 (2014)
  [arXiv:1404.7746 [hep-ph]].

\bibitem{Takahashi:2000te} 
  T.~T.~Takahashi, H.~Matsufuru, Y.~Nemoto and H.~Suganuma,
  Phys.\ Rev.\ Lett.\  {\bf 86}, 18 (2001)
  [hep-lat/0006005].

\bibitem{Takahashi:2002bw} 
  T.~T.~Takahashi, H.~Suganuma, Y.~Nemoto and H.~Matsufuru,
  Phys.\ Rev.\ D {\bf 65}, 114509 (2002)
  [hep-lat/0204011].

\bibitem{Hayano:2008vn} 
  R.~S.~Hayano and T.~Hatsuda,
  Rev.\ Mod.\ Phys.\  {\bf 82}, 2949 (2010)
  [arXiv:0812.1702 [nucl-ex]].

\end{thebibliography}
\end{document}